%% file: main.tex
\documentclass{article}

\input{_packages_arxiv}

\input{shortcuts}

\newcommand{\ABSTRACT}[1]{#1}

\title{Robust and efficient multiple-unit switchback experimentation}
\input{authors_arxiv}
\date{\today}

\begin{document}

\maketitle

\begin{abstract}
\input{chapters/0_abstract}
\end{abstract}

\input{chapters/1_intro}

\input{chapters/2_literature_review}
\input{chapters/3_setup}
\input{chapters/4_designs}

\input{chapters/5_simulations}
\input{chapters/6_experiments}
\input{chapters/7_conclusions}
\newpage
\bibliography{biblio}
\newpage
\appendix
\input{chapters/9_appendix}

\end{document}

%% file: _packages_arxiv.tex
\usepackage[margin=1in]{geometry}
\usepackage{amsmath}
\usepackage{amssymb}
\usepackage{amsthm}

\newtheorem{assumption}{Assumption}
\newtheorem{theorem}{Theorem}
\newtheorem{lemma}{Lemma}

\newtheorem{definition}{Definition}
\newtheorem{remark}{Remark}
\usepackage{bbm}
\usepackage{graphicx}
\graphicspath{{figures/}}
\usepackage{setspace}
\usepackage{natbib}
\usepackage{booktabs}
\usepackage{caption}
\usepackage{subcaption}
\usepackage{nicematrix}
\usepackage{multirow}

\usepackage{graphicx}
\graphicspath{{figures/}}
\usepackage{tikz}
\usepackage{pgfplots}
\pgfplotsset{compat=1.18}
\usepgfplotslibrary{statistics}  

\usepackage{hyperref}
\usepackage{cleveref}

\crefname{assumption}{Assumption}{Assumptions}
\Crefname{assumption}{Assumption}{Assumptions}

\crefname{theorem}{Theorem}{Theorems}
\Crefname{theorem}{Theorem}{Theorems}
\crefname{lemma}{Lemma}{Lemmas}
\Crefname{lemma}{Lemma}{Lemmas}
\crefname{proposition}{Proposition}{Propositions}
\Crefname{proposition}{Proposition}{Propositions}
\crefname{corollary}{Corollary}{Corollaries}
\Crefname{corollary}{Corollary}{Corollaries}
\crefname{definition}{Definition}{Definitions}
\Crefname{definition}{Definition}{Definitions}

%% file: shortcuts.tex
\newcommand{\ASIN}{{item}}
\newcommand{\ASINs}{{items}}
\newcommand{\AASIN}{{Item}}

\newcommand{\RR}{\mathbb{R}}
\newcommand{\EE}{\mathbb{E}}
\newcommand{\steptotal}{S}
\newcommand{\stepidx}{s}
\newcommand{\unittotal}{N}
\newcommand{\unitidx}{n}
\newcommand{\T}{\mathrm{T}}
\newcommand{\ynt}{Y_{\unitidx,\stepidx}^{\T}}
\newcommand{\C}{\mathrm{C}}
\newcommand{\ync}{Y_{\unitidx,\stepidx}^{\C}}
\newcommand{\lag}{\ell}

%% file: authors_arxiv.tex
\author{%
Paul Missault\thanks{Amazon, Luxemburg 1855, LU. Email: pmissaul@amazon.com}
\;
Lorenzo Masoero\thanks{Amazon, Seattle WA 98109. Email: masoerl@amazon.com}
\;
Christian Delbé\thanks{Amazon, Luxemburg 1855, LU. Email: delbec@amazon.com}
\;
Thomas Richardson\thanks{University of Washington, Seattle WA 98195, USA. Email: rchatho@amazon.com}
\;
Guido Imbens\thanks{Graduate School of Business and Department of Economics, Stanford University, SIEPR, NBER, Stanford, CA 94305, USA. Email: gimbens@amazon.com}
}

%% file: chapters/0_abstract.tex

\ABSTRACT{
	User-randomized A/B testing has emerged as the gold standard for online experimentation.
	However, when this kind of approach is not feasible due to legal, ethical or practical considerations, 
	experimenters have to consider alternatives like \ASIN{}-randomization.
	\AASIN{}-randomization is often met with skepticism due to its poor empirical performance.
	To fill this gap, in this paper we introduce a novel and rich class of experimental designs, ``Regular Balanced Switchback Designs'' (RBSDs).
	At their core, RBSDs work by randomly changing treatment assignments over both time and \ASINs{}.
	After establishing the properties of our designs in a potential outcomes framework, 
	characterizing assumptions and conditions under which corresponding estimators are resilient to the presence of carryover effects, 
	we show empirically via both realistic simulations and real e-commerce data that RBSDs systematically outperform
	standard \ASIN{}-randomized and non-balanced switchback approaches by yielding much more accurate estimates of the causal effects of interest without incurring any additional bias.
}

%% file: chapters/1_intro.tex

\section{Introduction} \label{sec:intro}

Online services have adopted A/B tests as the gold standard practice to enable data-driven decision-making and innovation.
Via randomization, A/B tests assess the effectiveness of an intervention on a metric of interest, relative to a baseline.
In these contexts, user-randomization, in which users interacting with the online service are randomly exposed to the intervention or the baseline, is typically considered the gold standard \citep{kohavi2020trustworthy, gupta2019top}.

In certain settings, however, a user-randomized experiment is either undesirable or unfeasible. 
For instance, user-level randomization cannot be used to test pricing changes on an e-commerce service, as it could constitute price discrimination, raising legal and ethical concerns \citep{Cooprider2023}. 
Additionally, user-level randomization can create complications in environments in which multiple systems with complex interaction patterns coexist. 
To illustrate this point, consider an e-commerce service testing the impact of using different \ASIN{} display names. 
For a given \ASIN{}, the control might render ``Stainless steel water bottle'' and the treatment  ``Vacuum Insulated Water Bottle for outdoor adventures''. 
User randomization allows measuring the \emph{direct} response of users due to the change of the display name, but fails to capture the potentially large \emph{indirect} effect of treatment through downstream systems.
Concretely, if the treatment affects how a search engine would rank the \ASIN{}, this will in turn impact metrics of interest measured at the user level.
Even when following best practices, estimates from a user-randomized experiment will not reliably capture these indirect effects.
At its core this is because a user-randomized experiment creates multiple co-existing versions of the same \ASIN{} within the service, creating significant risks for  downstream systems such as external search engines.
See \citet{google2025testing} for a discussion on how Google instructs website owners to set up their experiments.

To mitigate these ethical, legal and practical challenges, experimenters can set up experiments where the unit of randomization differs from the user. 
The natural choice is to randomize at the \ASIN{} level: when doing so, all consumers (including external search engines) have access to a single version of the \ASIN{}-level data.
Therefore,  \ASIN{}-randomization allows to naturally side step the \ASIN{}-multiplicity risk mentioned above, capturing both direct and indirect effects of treatment since downstream services like search engines can consume treated \ASINs{} and update their ranking.

While \ASIN{}-randomization can solve the aforementioned issues, it is often met with skepticism by experimenters.
Indeed, it is typically the case that \ASIN{}-randomized experiments lead to relatively lower power on experiments than user-randomization. This can be due to smaller inventory sizes relative to the customer base or due to large skew on the distribution of metrics of interest at the \ASIN{}-level.
To overcome these drawbacks of \ASIN{}-randomized experiments, we propose a novel experimental design which jointly randomizes across both \ASINs{} and time in a balanced way.
More precisely, we combine recent advances from time-randomized ``switchback'' experiments \citep{bojinov2023design}  with ideas from the literature on ``multi-unit'' online experimentation in the presence of interference \citep{masoero2023efficient, masoero2024multiple}.
Our novel class of experimental designs --- ``\textit{regular balanced switchback designs}'' (RBSD) --- allows the treatment assignment to vary randomly across both \ASINs{} and time-steps, but limits the treatment assignments such that they are doubly-balanced: RBSDs guarantee that (i) at every time step, a fixed fraction of \ASINs{} are treated and (ii) all \ASINs{} are assigned to treatment the same number of times across timesteps.

\begin{figure*}[ht]
\centering
    \centering
    \input{figures/tau_1_std_boxplot_intro}
    \caption{
    Distribution of the standard error of the average treatment effect for competing designs in \ASIN{}-randomized experiments. 
    See \cref{sec:experimental_designs} and \cref{sec:aa} for details on the designs and the experimental setup. 
    }
  \label{fig:intro_std_tau_1}
\end{figure*}
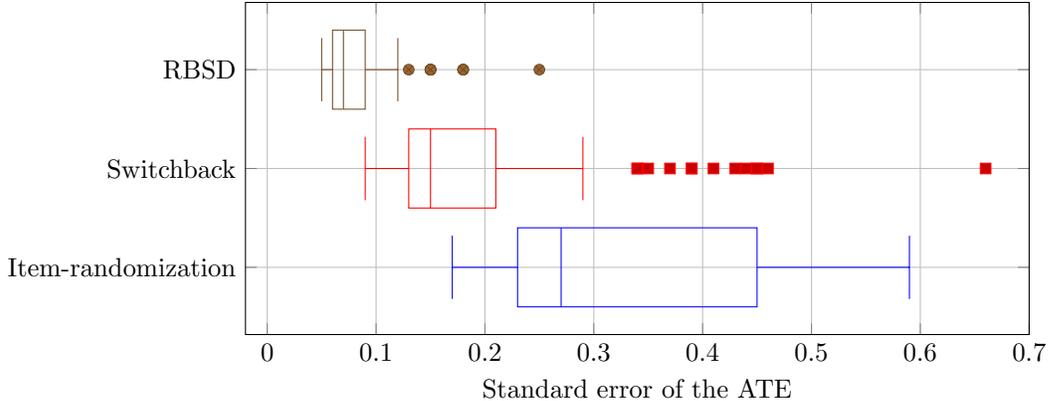

After reviewing relevant literature in \cref{sec:literature}, 
we provide the technical foundations of our contribution in \cref{sec:setup}, where we also explicitly characterize sufficient conditions under which we can build unbiased Horvitz-Thompson estimators of the \textit{average causal effect} under the potential outcome framework, even in the presence of carryover effects (or interference across time). 
We empirically validate our approach in \cref{sec:simulations,sec:exp_real}, where we show that ``doubly balancing'' allows to substantially decrease the variability of standard causal estimates of interest when compared to either a simple \ASIN{}-experiment, or a simple switchback experiment randomizing independently over time and \ASINs{}.
While further details are provided in \cref{sec:experimental_designs} and \cref{sec:aa}, \cref{fig:intro_std_tau_1} provides an intuition of this phenomenon: RBSDs yield much more precise estimates of the Average treatment effect; in turn this allows to detect the same effect with fewer datapoints, or -- alternatively -- a smaller effect when this is present, without inflating the rate of false discoveries.
We conclude in \cref{sec:conclusions} with a discussion of limitations of our method and future research directions.

%% file: figures/tau_1_std_boxplot_intro.tex
\begin{tikzpicture}
  \begin{axis}
    [
    xmin=-0.02,
    width=12cm,  
    height=6cm,  
    xmax=0.7,
    ytick={1,2,3},
    xlabel={Standard error of the ATE},
    yticklabels={\AASIN{}-randomization, Switchback, RBSD},
    yticklabel style={align=right, text width=3.5cm},
    boxplot/draw direction=x,
    grid=major,
    ]
    \addplot+[
    boxplot prepared={
        median=0.27,
        upper quartile=0.45,
        lower quartile=0.23,
        upper whisker=0.59,
        lower whisker=0.17,
      draw position=1
    },
    ] coordinates {(3,1.15) (3,0.97) (3,0.99) (3,1.16) (3,0.97) (3,0.95) (3,0.98) (3,1.55) (3,0.96) (3,1.02) (3,0.96) (3,1.38) (3,1.49) (3,1.40) (3,0.96) (3,1.22)};
    \addplot+[
    boxplot prepared={ 
    median=0.15,
    upper quartile=0.21,
    lower quartile=0.13,
    upper whisker=0.29,
    lower whisker=0.09,
      draw position=2
    },
    ] coordinates {(2,0.66) (2,0.46) (2,0.45) (2,0.34) (2,0.37) (2,0.39) (2,1.28) (2,0.34) (2,0.43) (2,0.73) (2,0.78) (2,0.45) (2,0.44) (2,2.09) (2,0.39) (2,0.41) (2,0.88) (2,0.35)}; 
    \addplot+[
    boxplot prepared={
    median=0.07,
    upper quartile=0.09,
    lower quartile=0.06,
    upper whisker=0.12,
    lower whisker=0.05,
      draw position=3
    },
    ] coordinates {(1,0.15) (1,0.13) (1,0.15) (1,0.18) (1,0.15) (1,0.18) (1,0.25) (1,0.15) (1,0.15) (1,0.18)}; 
  \end{axis}
  \path ([xshift=3cm]current axis.east) coordinate (dummy);
\end{tikzpicture}

%% file: chapters/2_literature_review.tex

\section{Summary of our contribution and related literature}
\label{sec:literature}


Randomized experiments where treatment exposure changes over time date back almost two centuries, rooted in agricultural applications where experimenters tried to optimize crop yields \citep{lawes1864report}.
These designs, together with Latin squares, were later formally introduced in the seminal work of \citet{fisher1935design}.
Soon after, their adoption became standard across a number of disciplines, and in particular in medical applications \citep[Chapter 1, Section 4]{jones2003design}.
Particularly relevant for the present discussion is the early work of \citet{cochran1939long}, who first formally discusses the risk of ``carryover'' effects in time randomized experiments --- namely, the fact that future outcomes for a given unit can depend on previous treatment assignments.
Later, \citet{cochran1941double} introduces estimation methods to detect both direct and carry-over effects via two orthogonal, \emph{balanced} Latin square designs.
Balancedness is achieved when each treatment occurs (i) the same number of times in each period and (ii) the same number of times for each subject. 
Moreover, the number of subjects who receive treatment $i$ in some period followed by treatment $j$ in the next period is the same for all $i \neq j$ \citep[Chapter 4]{jones2003design}. 
Building on early observation of \citet{yates1938gain}, \citet{williams1949experimental, williams1950experimental} later showed that balancedness guarantees improved efficiency of the corresponding estimators with respect to non-balanced alternatives. 


In the present work, we leverage insights from this classical literature to develop a practical design for online \ASIN{}-randomized experiments within the flexible framework of multiple randomized designs [MRDs] \citep{masoero2024multiple},
a large class of randomized experiments that, by jointly randomizing across multiple dimensions (\ASINs{}, timesteps, users, etc.), allow for the detection of spillover effects. 
Recently \citet{masoero2023efficient} showed that carefully designed MRDs randomizing jointly across \ASINs{} and time-steps, can yield unbiased estimators that are provably more efficient than those obtained via simple randomization.
However, their approach focuses on settings with no cross-unit or temporal interference, limiting its applicability in many real-world scenarios.

To address this limitation, we propose in \cref{sec:RBSDs} ``Regular Balanced Switchback Designs'' (RBSDs). 
These designs combine the provable efficiency gains and consequent power improvement of balanced multi-unit switchback experiments with robustness to carryover effects. 
We provide a theoretical framework and practical guidelines for implementation, extending the applicability of multiple-randomized designs to settings with potential temporal interference.
Our work can be viewed as a synthesis of two recent advancements in experimental design. 
On one hand, we build upon the efficiency gains demonstrated by \citet{masoero2023efficient}. 
On the other, we incorporate the robustness to carryover effects by following the approach of \citet{bojinov2023design}. 
This work proposes ``regular'' switchback designs that allow for unbiased estimation of treatment effects under assumptions about the temporal horizon of carryovers. 
While their approach is limited to single-unit implementations, our RBSDs extend this concept to multi-unit settings, resulting in a design that is both statistically efficient and resilient to temporal interference.


Next to these foundational papers, switchback experiments have also recently seen a surge in interest in the literature, particularly motivated by online applications.
See \citet{johnson2017online, liu2021trustworthy, hu2022switchback, ye2023deep, ni2023design, xiong2024data, xiong2024optimal} for a non-exhaustive list.
While our work is mainly motivated by improving power, and efficiency of \ASIN{}-randomized experiments while maintaining robustness to carryover effects, much of the recent literature on switchback designs has focused on mitigating bias introduced by interference across treated units. 
For instance, \citet{blake2014marketplace, holtz2024reducing} extensively discuss interference in online experiments, including temporal interference, and propose methods for detecting and quantifying the effects in these settings. 
Other researchers, such as \citet{li2022interference, doordash2019}, focus specifically on addressing carryover effects in switchback designs. The recent work by \citet{jia2025clusteredswitchbackdesignsexperimentation} is closely related to our work since that work also generalizes switchback to a multi-unit setting, and address simultaneously the carryovers and cross-unit spillovers. 
Our work complements these efforts by not only maintaining robustness to carryovers but also significantly enhancing the statistical power of the experimental design, thus addressing a gap in the existing literature.


Robustness to carryover effects in experimental designs has recently received considerable traction in the literature. While our work is closest to \citet{bojinov2023design}, two main approaches have emerged. We summarize these next.
The first approach --- ``interference networks'' \citep{eckles2017design, aronow2017sutva} --- involves modeling interference through networks. 
This method allows for unbiased estimates under arbitrary interference patterns, including spillovers that are both cross-\ASINs{} and over time (carryovers). 
While powerful, this approach assumes that the underlying interaction pattern causing carryover effects is known and correctly modeled as a network, which can be challenging in practice. 
The second approach, exemplified by the work of \citet{bojinov2023design}, does not require knowledge of an underlying network, and makes instead assumptions about the temporal horizon of carryover effects.
This method only requires specifying how far into the future a current treatment exposure can affect outcomes, which is often much simpler to determine based on domain knowledge. We note that this approach is subsumed by the interference network approach under careful specifications of the underlying network, as proved in Appendix~\ref{apdx:bojinov_aranov}.
Our work builds on this second approach, extending it to multi-unit settings while preserving its practical appeal and theoretical guarantees. 
By doing so, we strike a balance between the need for robust causal inference and the practical constraints of real-world experimentation.

%% file: chapters/3_setup.tex

\section{Experimenting with units over time} \label{sec:setup}

We now lay the technical foundations of our contribution. 
After introducing relevant notation in \cref{sec:notation} we discuss in \cref{sec:assumptions} assumptions on the data generating process under which standard Horvitz-Thompson estimators 
for the average treatment effect stemming from our designs are provably unbiased, as stated in \cref{sec:estimands_estimators}.
Proofs of our results can be found in Appendix~\ref{sec:proofs}.

\subsection{Notation} \label{sec:notation}

We consider an experiment with two treatments that runs for $\steptotal$ timesteps indexed by $\stepidx = 1,\ldots,\steptotal$ on $\unittotal$ eligible \ASINs{} indexed by $\unitidx  = 1,..., \unittotal$. 
Extensions to three or more treatments are straightforward, but omitted for simplicity in our discussion.
We let $W \in \{0,1\}^{\unittotal\times \steptotal}$ be a  matrix denoting treatment assignments, such that $W_{\unitidx,\stepidx}$ is an indicator for the treatment of unit $ \unitidx \in[\unittotal]~:=~\{1,\ldots, \unittotal\}$ at timestep $\stepidx \in [\steptotal]:=\{1,\ldots,\steptotal\}$; 
i.e., $W_{\unitidx, \stepidx}=1$ if \ASIN{} $n$ is in the treatment group at timestep $\stepidx$. 
We assume that there exists a ``potential outcome'' $Y_{\unitidx, \stepidx}(W)$, which is the value of a given metric of interest (e.g., clicks) for \ASIN{} $\unitidx$ at time $\stepidx$ when the randomization $W$ is applied.

Additionally, we let $\ync$ and $\ynt$ denote the potential outcome of \ASIN{} $n$ at timestep $\stepidx$  observed if all \ASINs{} were respectively never treated at any timestep, and always treated across all timesteps: 
\[
    \ync=Y_{\unitidx, \stepidx}
        \left(
        \mathbf{0}^{\unittotal \times \steptotal}
        \right)
        \quad 
            \text{and}
        \quad
    \ynt=Y_{\unitidx, \stepidx}
    \left(
        \mathbf{1}^{\unittotal \times \steptotal}
    \right),
\]
where $\mathbf{j}^{\unittotal \times \steptotal}$ denotes a matrix with all values $j$ of size ${\unittotal \times \steptotal}$, for $j\in \{0,1\}$.

\subsection{Assumptions on the data generating process} \label{sec:assumptions}

We now introduce useful assumptions on which we will rely to derive theoretical properties of our estimators. 
In what follows, $W_{\unitidx, \bullet}:=[W_{\unitidx,1}, \ldots, W_{\unitidx, \steptotal}] \in \{0,1\}^{1 \times \steptotal}$ denotes a row-vector encoding the \ASIN{} assignment over time.
 $W_{\bullet, \stepidx} = [W_{1,\stepidx}, \ldots, W_{\unittotal, \stepidx}]^\top \in \{0,1\}^{\unittotal\times1}$ denotes a column-vector encoding the assignments of all $\unittotal$ \ASINs{} at timestep $\stepidx$. 
For $\unitidx_2 > \unitidx_1$ and $\stepidx_2 > \stepidx_1$, $W_{\unitidx_1:\unitidx_2, \stepidx_1:\stepidx_2} \in \{0,1\}^{\unitidx_2-\unitidx_1, \stepidx_2 - \stepidx_1}$ denotes the ``sliced'' assignment submatrix which collects assignments $W_{\unitidx, \stepidx}$ for $n \in \{\unitidx_1,\ldots, \unitidx_2\}$ and $\stepidx \in \{\stepidx_1, \ldots, \stepidx_2\}$. 
Moreover, $W_{\unitidx_1:\unitidx_2, \bullet}$ is equivalent to $W_{\unitidx_1:\unitidx_2, 1:\steptotal}$ and $W_{\bullet, \stepidx_1:\stepidx_2}$ is equivalent to $W_{1:\unittotal, \stepidx_1:\stepidx_2}$.

\begin{assumption}[No cross-\ASIN{} spillovers]\label{as:sutva}
    The potential outcome for one \ASIN{} is a function of its own treatment assignments. Formally, for any $\unitidx \in [\unittotal]$ and $\stepidx \in [\steptotal]$, and any arbitrary sub-matrix assignments $W'_{1:\unitidx-1,\bullet}, W'''_{1:n-1,\bullet} \in \{0, 1\}^{\unitidx-1 \times \steptotal}$, $W''_{\unitidx+1:\unittotal,\bullet},W''''_{\unitidx+1:\unittotal,\bullet} \in \{0, 1\}^{\unittotal-
    \unitidx \times \steptotal}$, it holds:
    \begin{equation} \label{eq:SUTVA}
        Y_{\unitidx, \stepidx}
        \left(
            \begin{bmatrix}
                W'_{1:\unitidx-1,\bullet} \\
                W_{\unitidx,\bullet} \\
                W''_{\unitidx+1:\unittotal,\bullet}
                \end{bmatrix} 
            \right) 
            = 
        Y_{\unitidx, \stepidx}
        \left(
            \begin{bmatrix}
                W'''_{1:\unitidx-1,\bullet} \\
                W_{\unitidx,\bullet} \\
                W''''_{\unitidx+1:\unittotal,\bullet} 
            \end{bmatrix}
        \right)
        = 
        Y_{\unitidx, \stepidx}
        \left(
            W_{\unitidx,\bullet}
        \right).
    \end{equation} 
\end{assumption}

\Cref{as:sutva} is crucial  for all randomized controlled trials, and is often referred to as the stable-unit-treatment-value assumption, or SUTVA \citep{imbens2010rubin}. 
In practice, SUTVA is often violated due to the presence of interference across units (see, e.g., \citet{bajari2023experimental} for a review). 
Interference can arise in any experiment, including one where the unit of randomization is a user, due to ``network effects'' \citep{saveski2017detecting, saint2019using}.
In \ASIN{}-randomized experiments, interference can occur when \ASINs{} are in direct competition with each other (e.g., substitute goods in a pricing experiment).
All our results rely on \cref{as:sutva}. 

\begin{remark}
\label{remark:sutva}
When experimenters are concerned about violations of \cref{as:sutva}, the standard approach is to form clusters of \ASINs{} ``blocking'' interference effects (e.g., competing goods), and randomize at the \ASIN{}-cluster level \citep{hudgens2008toward, gerber2012field}. Recent work by \citet{jia2025clusteredswitchbackdesignsexperimentation} discusses clustering specifically for switchback experimentation. 
We discuss limitations and extensions of our approach in \cref{sec:conclusions}.
\end{remark}

\begin{assumption}[Non-anticipating Potential Outcomes] \label{as:potential}  
Let \cref{as:sutva} hold.
Additionally, for any $\unitidx \in [\unittotal]$, $\stepidx \in [\steptotal]$, and  $W_{\bullet,1:\stepidx} \in \{0, 1\}^{\unittotal \times \stepidx}$,  $W'_{\bullet, \stepidx+1:\steptotal},W''_{\bullet, \stepidx+1:\steptotal} \in \{0, 1\}^{\unittotal \times \steptotal - \stepidx}$, it holds
    \[
        Y_{\unitidx, \stepidx}\left(W_{\unitidx,1:\stepidx}, W'_{\unitidx,\stepidx+1:\steptotal}\right) 
        = 
        Y_{\unitidx, \stepidx}\left(W_{\unitidx,1:\stepidx}, W''_{\unitidx,\stepidx+1:\steptotal}\right).
    \]
\end{assumption}

Under \cref{as:potential}, potential outcomes at a given timestep do not depend on \textit{future} assignments.

\begin{assumption}[No Carryover Effects] \label{as:no_carryover} 
Let \cref{as:sutva} hold.  
For any $\unitidx \in [\unittotal]$, $\stepidx \in \{2, \ldots , \steptotal\}$, $W_{\bullet, \stepidx:\steptotal}\in\{0,1\}^{\unittotal \times \steptotal-\stepidx+1}$, and $W'_{\bullet,1:\stepidx-1},W''_{\bullet,1:\stepidx-1} \in \{0, 1\}^{\unittotal \times \stepidx-1}$,
\[
    Y_{\unitidx, \stepidx}(W'_{\unitidx,1:\stepidx-1}, 
    W_{\unitidx, \stepidx:\steptotal}) 
    = Y_{\unitidx, \stepidx}(W''_{\unitidx,1: \stepidx-1}, W_{\unitidx, \stepidx:\steptotal}).
\]
\end{assumption}

Under \cref{as:no_carryover}, potential outcomes at a given timestep do not depend on \textit{past} assignments.
This can be very restrictive in time-randomized designs, with concerns being both behavioral (e.g., users might need time to "learn" or can memorize past treatments) and technical (e.g., systems latency in rendering the treatment).
In subsequent sections, we will discuss how to relax this assumption and propose an estimator of causal effects that is unbiased even in the presence of carryover effects (of known duration).

\begin{assumption}[$m$-Carryover Effects]\label{as:carryover} Let \cref{as:sutva} hold.  There exists a fixed and given $m \in \{1,2,\ldots,\steptotal-1\}$, such that for any $\unitidx \in [\unittotal]$, $\stepidx\in\{m+1, \dots , \steptotal\}, W_{\bullet, \stepidx-m:\steptotal}\in\{0,1\}^{\unittotal \times \steptotal-\stepidx+m+1}$, and for any $W'_{ \bullet,1:\stepidx-m+1},W''_{\bullet,1:\stepidx-m+1} \in \{0, 1\}^{\unittotal \times \stepidx-m+1}$, the following holds:
$$
    Y_{\unitidx, \stepidx}(W'_{n,1:\stepidx-m-1}, W_{\unitidx, \stepidx-m:\steptotal}) 
    = 
    Y_{\unitidx, \stepidx}(W''_{\unitidx,1:\stepidx-m-1}, W_{\unitidx,\stepidx-m:\steptotal})
$$
\end{assumption}
Under \cref{as:carryover}, the outcome observed for unit $\unitidx$ at the $\stepidx$-th timestep, depends at most on $m$ previous assignments.

\subsection{Estimands and estimators for the average treatment effects} \label{sec:estimands_estimators}

Our estimand of interest is the average treatment effect \citep{imbens2010rubin}:
\begin{equation}
\label{eq:tau}
    \tau=\frac{1}{\unittotal\steptotal}\sum_{n=1}^{\unittotal}\sum_{\stepidx = 1}^{\steptotal} \left\{ \ynt - \ync \right\}.
\end{equation}
$\ynt$ and $\ync$ cannot both be observed: each pair $(\unitidx, \stepidx)$ is exposed to either treatment or control, not to both. 
In a randomized experiment, we expose (\ASIN{}-time) pairs to different treatments encoding their assignments via $W \in \{0,1\}^{\unittotal \times \steptotal}$. 
We then \emph{estimate} the effect of interest $\tau$ introduced in \Cref{eq:tau} from data with the ``Horvitz-Thompson'' estimator $\hat{\tau}(W)$ \citep{horvitz1952generalization}: 
\begin{align} \label{eq:hv_th}
\begin{split}
    \hat{\tau}(W) = 
        \frac{1}{\unittotal\steptotal} 
            \sum_{\unitidx = 1}^{\unittotal} 
            \sum_{\stepidx = 1}^{\steptotal}  & 
                \left\{ \frac{W_{\unitidx, \stepidx}}{P(W_{\unitidx, \stepidx}=1)}Y_{\unitidx, \stepidx}(W)  \right.\\
        & \left. - \frac{(1-W_{\unitidx, \stepidx})}{P(W_{\unitidx, \stepidx}=0)}Y_{\unitidx, \stepidx}(W) \right\}
\end{split}
\end{align}
Here, $P(\cdot)$ denotes the probability of an assignment, which is typically known (and can be enforced) by the experimenter in the context of a randomized control trial (RCT, or A/B test).

Carryover effects, where the outcome measured at a given timestep $\stepidx$ depends on the unit's previous treatment assignments, are a typical risk in switchback experiments. 
When these effects are present, the average treatment effect of \cref{eq:tau} itself is ill-defined. Consider how we would need information or assumptions about the pre-experiment period since observations at $\stepidx =1$ are impacted by pre-experiment assignments (at $\stepidx \leq 0$). 
To allow for a well-defined average treatment effect, without assumptions on the pre-experiment period, 
we consider an alternative estimand $\tau_\lag$, obtained by discarding a pre-defined number of ``burn-in'' $\lag \ge 0$ timesteps occurring at the beginning of the experiment.
The intuition is that when a carryover effect is bounded by $\lag$ (i.e., an outcome measured at a timestep can only be impacted by at most by  $\lag$ previous timesteps), the new estimand will be well defined.
We define the corresponding \textit{lag-$\lag$} average causal effect $\tau_\lag$ in \cref{eq:lag_p_tau}.
This definition mirrors that of \citet{bojinov2023design} but is adapted to the multi-unit setting:
\begin{equation} \label{eq:lag_p_tau}
\tau_\lag = 
    \frac{1}{\unittotal(\steptotal-\lag)}
    \sum_{\unitidx=1}^{\unittotal}
    \sum^{\steptotal}_{\stepidx = \lag+1}
        ( \ynt - \ync ).
\end{equation}

To estimate the \textit{lag-$\lag$} causal effect, we propose the following estimator: 

\begin{align}\label{eq:lag_p_hv_th}
\begin{split}
    \hat{\tau}_\lag(W) & = 
        \frac
            {1}
            {\unittotal(\steptotal-\lag)} 
        \sum_{\unitidx=1}^{\unittotal} 
        \sum_{\stepidx = \lag+1}^{\steptotal} 
            Y_{\unitidx, \stepidx}(W) 
        \\
        & 
        \left[ 
            \frac
                {1(W_{\unitidx,\stepidx-\lag:\stepidx}=\mathbf{1}^{\lag+1})}
                {P(W_{\unitidx,\stepidx-\lag:\stepidx}=\mathbf{1}^{\lag+1})}
            -
              \frac
                {1(W_{\unitidx,\stepidx-\lag:\stepidx}=\mathbf{0}^{\lag+1})}
                {P(W_{\unitidx,\stepidx-\lag:\stepidx}=\mathbf{0}^{\lag+1})}
        \right]. 
\end{split}
\end{align}

We conclude this section by formally characterizing assumptions under which $\hat{\tau}$ is an unbiased estimator of ${\tau}$ and $\hat{\tau}_\lag$ is an unbiased estimator of ${\tau}_\lag$, and provide corresponding variance estimators.

\begin{theorem}[Unbiasedness and variance estimation of $\hat{\tau}$ for $\tau$ under Assumptions \ref{as:sutva}, \ref{as:potential} and \ref{as:no_carryover}] \label{thm:unbiased}
    Consider a randomized experiment with two treatments, $\unittotal$ \ASINs{} and $\steptotal$ timesteps. 
    For any randomization design over both \ASINs{} and timesteps and such that $P(W_{\unitidx, \stepidx}~=~1)~ \in~ (0,1)$ for all $n$ and $\stepidx$, \cref{eq:hv_th} provides an unbiased estimate of \cref{eq:tau} when \cref{as:sutva}, \ref{as:potential} and \ref{as:no_carryover} hold.
    Moreover, defining the individual treatment effect ($\widehat{\mathrm{ITE}}_\unitidx$) as
    $$
    \widehat{\mathrm{ITE}}_\unitidx(W) = 
    \frac{1}{\steptotal} 
    \sum_{\stepidx = 1}^{\steptotal} 
    \left[ 
        \frac
            {W_{\unitidx, \stepidx}}
            {P(W_{\unitidx, \stepidx}=1)}
        Y_{\unitidx, \stepidx}(W) - 
        \frac
            {(1-W_{\unitidx, \stepidx})}
            {P(W_{\unitidx, \stepidx}=0)}
        Y_{\unitidx, \stepidx}(W) 
    \right],
    $$
    a consistent estimator of the standard error of $\hat{\tau}$ is given by:
    $$
    \hat{\sigma}(W)
    :=
    \sqrt{
    \frac
        {1}
        {\unittotal(\unittotal-1)} 
    \sum_{\unitidx=1}^{\unittotal} 
        \left\{
            \widehat{\mathrm{ITE}}_\unitidx(W) - \hat{\tau}(W)
        \right\}^2
    }.
    $$
\end{theorem}

\begin{theorem}[Unbiasedness and variance estimation of $\hat{\tau}_\lag$ for $\tau_\lag$ under Assumptions \ref{as:sutva}, \ref{as:potential} and \ref{as:carryover}] \label{thm:unbiased_lagp}
    Consider a randomized experiment with two treatments, $\unittotal$ \ASINs{} and $\steptotal$ timesteps. 
    Under any randomization design over both \ASINs{} and timesteps and such that $P(W_{\unitidx, \stepidx}=1) \in (0,1)$ for all $n$ and $\stepidx$, \cref{eq:lag_p_hv_th} is an unbiased estimate of \cref{eq:lag_p_tau} when \cref{as:sutva}, \ref{as:potential} and \ref{as:carryover} hold.
    Moreover, defining the $\lag$-lagged individual treatment effect ($\widehat{\mathrm{ITE}}_\unitidx^{(\lag)}$) as
    $$
    \widehat{\mathrm{ITE}}^{(\lag)}_\unitidx(W) = 
        \frac
            {1}
            {(\steptotal-\lag)} 
            \sum_{\stepidx = \lag+1}^{\steptotal} 
            Y_{\unitidx, \stepidx}(W) 
            \left[ 
            \frac
                {1(W_{\unitidx,\stepidx-\lag:\stepidx}=\mathbf{1}^{\lag+1})}
                {P(W_{\unitidx,\stepidx-\lag:\stepidx}=\mathbf{1}^{\lag+1})}
            -
          \frac
                {1(W_{\unitidx,\stepidx-\lag:\stepidx}=\mathbf{0}^{\lag+1})}
                {P(W_{\unitidx,\stepidx-\lag:\stepidx}=\mathbf{0}^{\lag+1})}
        \right],
    $$
    then, a consistent estimator of the standard error of $\hat{\tau}_\lag$ is given by:
    $$
    \hat{\sigma}_{\lag}(W)
    :=
    \sqrt{\frac
        {1}
        {\unittotal(\unittotal-1)} 
    \sum_{\unitidx=1}^{\unittotal} 
        \left\{
            \widehat{\text{ITE}}^{(\lag)}_\unitidx(W) - \hat{\tau}_\lag(W) .
        \right\}^2
    }.
    $$
\end{theorem}

The proof of \cref{thm:unbiased} can be found in Appendix \ref{sec:proof_unbiased}, \cref{thm:unbiased_lagp} in Appendix \ref{sec:proof_unbiased_lagp}.

%% file: chapters/4_designs.tex

\section{Experimental designs} \label{sec:experimental_designs}

In this section we formally introduce a number of experimental designs combining \ASIN{} and time randomization. 
In what follows, we call ``experimental design'' a probability distribution (or a class thereof) over a suitable space of treatment assignments (e.g., binary matrices with dimension $\unittotal \times \steptotal$).

\subsection{Designs}
\subsubsection{\AASIN{}-randomized designs} \label{sec:conventional} 

\AASIN{}-randomized designs consist of a single randomized treatment assignment at the start of the experiment that is left unaltered throughout the experiment duration.
\begin{definition}[\ASIN{}-randomization]
    Given $\unittotal$ items, and a parameter $p \in (0,1)$ such that $p \unittotal \in \mathbb{N}$, an \ASIN{}-randomized design is the probability distribution on $\{0,1\}^{\unittotal \times \steptotal}$ stemming from (i) choosing at random without replacement $p \unittotal$ of the total $\unittotal$ units and (ii) letting $Z_\unitidx$ identify whether $\unitidx$ was chosen, setting $W_{\unitidx, \stepidx} ~=~ Z_\unitidx$, $ \unitidx=1,\ldots,\unittotal$ and $\stepidx=1,\ldots,\steptotal$.
    Equivalently, an \ASIN{}-randomized design with parameters $p, \unittotal$ is the uniform distribution over the following space of assignment matrices: 
    $$
    	\mathfrak{W}^{\text{\ASIN{}}} 
            := 
            \big\{ W \in \{0,1\}^{\unittotal \times \steptotal}\; 
            :\; W_{\unitidx, \stepidx} = W_{\unitidx,1} 
            \forall \unitidx \in [\unittotal], \; \forall \stepidx \in [\steptotal]
            \land \sum_\unitidx W_{\unitidx, 1} = p\unittotal\big\}.
    $$
    \end{definition}

\Cref{unit_randomized} is a valid treatment assignment stemming from an \ASIN{}-randomized experiment with $\unittotal=5$ and $p = 0.6$. Units $\unitidx = 2,3,5$ are selected, and exposed  to treatment for $\stepidx = 1, \ldots, 6$:
\begin{equation}\label{unit_randomized}
\arraycolsep=10pt\def\arraystretch{0.65}
\begin{NiceArray}{rl@{\hspace{.1em}}cccccc@{\hspace{1em}}c}[%
name=superwmat,%
create-extra-nodes,%
extra-margin=0pt,%
cell-space-limits=3pt,%
baseline=7]
& & & & & & & & \unitidx \\
&\stepidx \rightarrow &1 & 2 & 3 & 4 & 5 & 6 & \rotate \leftarrow \\ 
& & 0 & 0 & 0 & 0 & 0 & 0 & 1 \\
& & 1 & 1 & 1 & 1 & 1 & 1 & 2 \\
& & 1 & 1 & 1 & 1 & 1 & 1 & 3 \\
& & 0 & 0 & 0 & 0 & 0 & 0 & 4 \\
& & 1 & 1 & 1 & 1 & 1 & 1 & 5
\CodeAfter
\SubMatrix({3-3}{7-8})[name=wmat]
\begin{tikzpicture}
\node[left] (W) at (wmat-left.west) {$W =$} ;
\end{tikzpicture}
\end{NiceArray}
\begin{tikzpicture}[remember picture,overlay]
\end{tikzpicture}
\end{equation}

\subsubsection{Time-randomized designs (``switchback'' experiments)} \label{sec:switchback}
``Switchback'' or time-randomized experiments, are ones where all \ASINs{} are randomly assigned to either treatment or control at every timestep $\stepidx$. 
In the most general case, the treatment assignment of all \ASINs{} might randomly change at every timestep $\stepidx$ without constraints:
\begin{equation}\label{switchback}
    \arraycolsep=10pt\def\arraystretch{0.65}
        \begin{NiceArray}{rl@{\hspace{.1em}}cccccc@{\hspace{1em}}c}[%
        name=superwmat2,%
        create-extra-nodes,%
        extra-margin=0pt,%
        cell-space-limits=3pt,%
        baseline=7]
        & & & & & & & & \unitidx \\
        & \stepidx \rightarrow &1 & 2 & 3 & 4 & 5 & 6 & \rotate \leftarrow \\ 
        & & 1 & 1 & 0 & 0 & 0 & 1 & 1 \\
        & & 1 & 1 & 0 & 0 & 0 & 1 & 2 \\
        & & 1 & 1 & 0 & 0 & 0 & 1 & 3 \\
        & & 1 & 1 & 0 & 0 & 0 & 1 & 4 \\
        & & 1 & 1 & 0 & 0 & 0 & 1 & 5
        \CodeAfter
        \SubMatrix({3-3}{7-8})[name=wmat]
        \begin{tikzpicture}
        \node[left] (W) at (wmat-left.west) {$W =$} ;
        \end{tikzpicture}
        \end{NiceArray}
        \begin{tikzpicture}[remember picture,overlay]
    \end{tikzpicture}
\end{equation}

\begin{definition}[Switchback-randomization]
    Given $\steptotal$ timesteps, and a parameter $p \in (0,1)$ such that $p \steptotal \in \mathbb{N}$, a switchback design is the probability distribution on $\{0,1\}^{\unittotal \times \steptotal}$ stemming from (i) choosing at random without replacement $p \steptotal$ of the total $\steptotal$ timesteps and (ii) letting $Z_\stepidx$ identify whether $\stepidx$ was chosen, setting $W_{\unitidx, \stepidx} ~=~ Z_\stepidx$, $ \stepidx=1,\ldots,\steptotal$ and $\unitidx=1,\ldots,\unittotal$.
    Equivalently, a switchback design with parameters $p, \steptotal$ is the uniform distribution over the space of assignment matrices: 
    $$
    	\mathfrak{W}^{\text{Switchback}} 
            := 
            \big\{ W \in \{0,1\}^{\unittotal \times \steptotal}\; 
            :\; W_{\unitidx, \stepidx} = W_{1, \stepidx} 
            \forall \stepidx \in [\steptotal], \; \forall \unitidx \in [\unittotal]
            \land \sum_\stepidx W_{1, \stepidx} = p\steptotal\big\}.
    $$
    \end{definition}

\subsubsection{Time and \ASIN{}-randomized designs (``Multi-Unit Switchback'' experiments)}

In a ``multi-unit switchback'' experiment, the experimenter has the ability to re-determine whether each \ASIN{} $\unitidx$ is assigned to either treatment or control at every timestep $\stepidx$.
In the most general case, the treatment assignment of each \ASIN{} $\unitidx$ might randomly change at every timestep $\stepidx$ without constraints:
\begin{equation}\label{switchback_multi}
    \arraycolsep=10pt\def\arraystretch{0.65}
        \begin{NiceArray}{rl@{\hspace{.1em}}cccccc@{\hspace{1em}}c}[%
        name=superwmat3,%
        create-extra-nodes,%
        extra-margin=0pt,%
        cell-space-limits=3pt,%
        baseline=7]
        & & & & & & & & \unitidx \\
        & \stepidx \rightarrow &1 & 2 & 3 & 4 & 5 & 6 & \rotate \leftarrow \\ 
        & & 0 & 1 & 0 & 1 & 1 & 1 & 1 \\
        & & 1 & 1 & 1 & 0 & 0 & 1 & 2 \\
        & & 1 & 0 & 1 & 0 & 0 & 1 & 3 \\
        & & 0 & 1 & 0 & 0 & 0 & 0 & 4 \\
        & & 1 & 1 & 0 & 1 & 0 & 1 & 5
        \CodeAfter
        \SubMatrix({3-3}{7-8})[name=wmat]
        \begin{tikzpicture}
        \node[left] (W) at (wmat-left.west) {$W =$} ;
        \end{tikzpicture}
        \end{NiceArray}
        \begin{tikzpicture}[remember picture,overlay]
    \end{tikzpicture}
\end{equation}

\begin{definition}[Multi-unit switchback designs]
    A probability distribution on $\{0,1\}^{\unittotal \times \steptotal}$ is a switchback design. E.g., the uniform distribution over $\{0,1\}^{\unittotal \times \steptotal}$ is a multi-unit switchback design.
\end{definition}

\subsection{Regularity and balancedness} \label{sec:regular_balanced}

A rich and long line of research has carefully characterized conditions under which switchback designs yield estimators that consistently recover treatment effects in the presence of carryover effects.
Most recently \citet{bojinov2023design,ni2023design} focused on ``regular'' designs, which we formally introduce in \cref{sec:regular}.
Secondly, it has long been observed that under certain \emph{balancing} conditions on the randomization scheme, multi-unit switchback designs provably increase the efficiency of the resulting estimators (see e.g.\ \citet[Chapter 4]{jones2003design} and the references therein). 
Recently, \citet{masoero2023efficient} characterized efficiency properties of multi-unit balanced designs in the absence of cross-unit interference. 
We formally introduce balanced designs in \cref{sec:balanced}.
We combine these two properties (regularity, and balancedness) and introduce in \cref{sec:RBSDs} multi-unit randomized designs that are both regular and balanced --- RBSDs. 
We later empirically show in \cref{sec:simulations,sec:exp_real} that RBSDs yield estimates of the causal effects of interest that are both robust to carryover effects and more efficient than competing methods.

\subsubsection{Regular switchback designs} \label{sec:regular}
\citet{bojinov2023design} extensively discuss time-randomized experiments robust to carryover effects. 
We here extend these designs to accommodate joint randomization over both timesteps \emph{and} \ASINs{}.


\begin{definition}[Regular switchback design]\label{def:regular} 
    Given any selection of $K<\steptotal$ breakpoints $\mathbb{K} ~=~ \{1:=\stepidx_0<\stepidx_1<...<\stepidx_K \le \steptotal \} \subseteq [\steptotal]$, 	and any collection $\mathbb{Q} ~=~ [q_0,q_1,...,q_K]^\top \in (0,1)^{K+1}$ of weights in $(0,1)$, a $(\mathbb{K}, \mathbb{Q})$-regular switchback design is a probability distribution $\mathbf{P}$ on the space of matrix assignments $\{0,1\}^{\unittotal \times \steptotal}$ such that for $\stepidx \in [\stepidx_k, \stepidx_{k+1})$ a supported assignment $W$ satisfies
    \begin{equation}
        \mathbf{P}(W_{\unitidx, \stepidx} = 1) = q_k 
    \quad
    \text{and} 
    \quad
        W_{\unitidx, \stepidx}  = W_{\unitidx, \stepidx_k}.
        \label{eq:regular_assignment}
    \end{equation}
\end{definition}
Despite its seemingly convoluted definition, sampling valid matrix assignments from a regular switchback design is trivial: for each unit $\unitidx$ we flip $K+1$ independent biased coins (with $q_k > 0$ and $q_k<1$), $W_{\unitidx, \stepidx_k} ~\sim~ \mathrm{Bernoulli} (q_k)$, $k=0,\ldots,K$.
Each treatment assignment is defined by the outcome of the corresponding coin via \cref{eq:regular_assignment}.
By construction, the treatment assignment of a unit does not change between two breakpoints of $\mathbb{K}$. I.e., if two timesteps $\stepidx < \stepidx'$ satisfy $(\stepidx, \stepidx')\subset [\stepidx_{k}, \stepidx_{j})$ for some $k, j$, then $W_{\unitidx, \stepidx} = W_{\unitidx, \stepidx'}$.
If $W \in \{0,1\}$ is sampled via the procedure above, every unit $\unitidx$ (i.e., a row in the assignment matrix $W$) is \textit{regular} in the sense of \citet{bojinov2023design}. 
Our definition also excludes any ``degenerate'' design where $W_{\unitidx, \stepidx}$ is guaranteed to have a pre-determined allocation (since that would imply $P(W_{\unitidx, \stepidx} = 1) \in \{0,1\}$), or where cyclical exposure patterns are enforced.   
Under the assumptions of \cref{thm:unbiased_lagp}, the  estimator $\hat{\tau}_\lag$ defined in \cref{eq:lag_p_hv_th} obtained from a regular switchback design is unbiased for the causal lag-$\lag$ estimand $\tau_\lag$ of \cref{eq:lag_p_tau}.

\begin{assumption}[Bounded potential outcome] \label{as:bounded}  
    The potential outcomes $Y_{n,s}(W)$ are bounded for any choice of $\unitidx$, $\stepidx$ and $W$, i.e., $\exists B > 0$ such that $\forall \;\stepidx \in [\steptotal], \forall \; \unitidx \in [\unittotal], \forall \; W \in \{0,1\}^ {\unittotal \times \steptotal}, \; |Y_{\unitidx, \stepidx}(W)| \leq B$.
\end{assumption}

\begin{lemma}
\label{lemma:optimal}
    Fix a number of breakpoints $K \in \mathbb{N}$. 
    Under assumptions \ref{as:sutva}, \ref{as:potential}, \ref{as:carryover} and \ref{as:bounded}, for a given carryover effect order $m$ as defined in \cref{as:carryover}, the minimax optimal choices of $\mathbb{K}^\star \in [S]^K$ and $\mathbb{Q}^\star \in [0,1]^K$ minimizing the variance of the estimator $\hat{\tau}_\lag$ in \cref{eq:lag_p_hv_th} under adversarial selection of the potential outcomes for an experiment over $\steptotal$ timesteps are respectively the solution to
    $$
        \mathbb{K}^\star=
        \min_{\mathbb{K} \subset[\steptotal]^K} 
        \left\{
            4\sum_{k=0}^K (s_{k+1}-s_k)^2 
            + 8m(s_K-s_1) + 4m^2K - 4m^2 
            + 4\sum_{k=1}^{K-1}[(m-s_{k+1}+s_k)^+]^2
        \right\},
    $$
    and $\mathbb{Q}^\star = \left[\frac{1}{2},\frac{1}{2}, \dots,\frac{1}{2}\right]^\top$, where $(a)^+:= \max\{0, a\}$.
\end{lemma}
See \cref{sec:proof_lemma_optimal} for a proof.
Notice that \cref{lemma:optimal} -- being a minimax result -- specifies the design  minimizing the variance of the estimator only in the worst-case, providing guarantees on the upper bound of the variance rather than minimizing the expected variance under ``typical'' conditions. 
In practice, on real data, different designs can achieve lower variance than the one suggested by \cref{lemma:optimal}.
In particular, specifying a balanced randomization enforcing equal numbers of treatment and control units when the sample size is small often yields lower variance in practice.

\subsubsection{Balanced switchback designs} \label{sec:balanced}
\Cref{thm:unbiased,thm:unbiased_lagp} provide conditions under which the regular design of \cref{def:regular} yield unbiased estimates for $\tau$ and $\tau_\lag$.
These estimators can have large variance as we will show in \cref{sec:aa}. For that reason, \citet{masoero2023efficient} proposed provably efficient designs for setting of multi-unit randomization  -- ``\textit{balanced switchback designs}''.

\begin{definition}[Balanced Switchback Designs]\label{def:balanced}
Given $\unittotal$, $\steptotal$, and a fixed probability of treatment $p \in (0,1)$ such that $pN, pS \in \mathbb{N}$, the uniform probability over the following space of assignment matrices is  a $p$-balanced switchback design:
$$
    \mathfrak{W}^{\mathrm{BSD}}_p := 
    \left\{ 
            W \in \{0,1\}^{\unittotal \times \steptotal}: 
    \forall n \in [N], \sum_{\stepidx=1}^\steptotal W_{\unitidx, \stepidx} = p\steptotal 
    \;  
    , 
    \forall\stepidx \in [S] \sum_{\unitidx=1}^\unittotal W_{\unitidx, \stepidx} = p\unittotal           
   \right\}.
$$
\end{definition}

\subsubsection{Regular balanced switchback designs [RBSDs]} \label{sec:RBSDs}
It is important to note that not every regular assignment matrix is balanced.
Consider, for example, an experiment with $\steptotal=10$ total timesteps, $\mathbb{K} = \{1,2,3\}$, and $\mathbb{Q} = [\frac{1}{2},\frac{1}{2},\frac{1}{2},\frac{1}{2}]^\top$.
This design satisfies \cref{eq:regular_assignment}, but it is not balanced.
Indeed, the treatment assignment defined at $s=3$ determines whether an \ASIN{} $\unitidx$ will be in treatment or control for the subsequent $7$ timesteps, and so it is not the case that a unit will be exposed to treatment and control the same number of timesteps. 
Conversely, we do note that balanced designs are automatically regular with $\mathbb{K} = [\steptotal]$ and $\mathbb{Q} = [p, \ldots, p] \in (0,1)^{K+1}$, but that this limits the experimenter in their choice of randomization points: balancedness as defined by \citet{masoero2023efficient} implies randomization happens at every timestep $\stepidx=1,\dots,\steptotal$, with a constant treatment assignment probability equal to $p$.

In order to keep the flexibility of choosing randomisation points we propose a design that samples from the space of assignment matrices that are \emph{both} regular \emph{and} balanced for an arbitrary choice of $K$ . Regularity ensures $\hat{\tau}_\lag$ is reliable estimator of $\tau_\lag$ even in presence of carryover effects, and balancedness ensures efficiency of the resulting estimators. 
\begin{definition}[Regular balanced switchback designs (RBSDs)]
    Given $\unittotal$ and $\steptotal$, let $p ~\in~ (0,1)$ be such that $p \unittotal, p \steptotal \in \mathbb{N}$. Let $\mathbb{K} =\{1=: \stepidx_0 < \stepidx_1 < \dots < \stepidx_K \le \steptotal \}$ be a collection of breakpoints and $\mathbb{Q} = [ p, p, \ldots, p ]^\top \in (0,1)^{K+1}$ a collection of weights.
    A regular balanced switchback design is a probability distribution over assignment matrices corresponding to the following sampling mechanism:
     (i) first sample from a $(\mathbb{K}, \mathbb{Q})$-regular design and (ii) retain the sample only if it is $p$-balanced, i.e.\ if it is an assignment in $ \mathfrak{W}^{\mathrm{BSD}}_p$ as per \cref{def:balanced}.
\end{definition}

Sampling a regular balanced switchback design is not trivial.
We provide in lemmas \ref{lemma:permutation} and \ref{lemma:complement} a way to efficiently sample from a RBSD  --- i.e., a practical algorithm to obtain a regular and balanced assignment $W$ for specific choices of $\mathbb{K}$ and $\mathbb{Q}$.

\begin{lemma}\label{lemma:permutation} 
    Let $p \in (0,1)$ be chosen such that $p \steptotal$ is integer valued, and let $\mathbf{x} \in \{0,1\}^\steptotal$ with $\|\mathbf{x}\|_1 = p \steptotal$. 
    The experimental design that samples matrix assignments $W \in \{0,1\}^{\unittotal \times \steptotal}$ by stacking vertically $\unittotal$ random permutations of the vector $\mathbf{x}$ is a \textit{regular} switchback design with $\mathbb{K}=\{1,2,..., \steptotal\}$ and $\mathbb{Q}= \{p, p ,.., p\}$ 
    whenever $1 < pS < S-1$.
\end{lemma}

\begin{lemma}\label{lemma:complement} 
Let $\unittotal$ be an even number. For an assignment matrix $\tilde{W} \in \{0, 1\} ^ {\frac{\unittotal}{2} \times \steptotal}$,
let
$\tilde{W}':=1^{\frac{\unittotal}{2} \times \steptotal} - \tilde{W}$ be its complement.
The matrix $W$ obtained by vertically stacking  $\tilde{W}$ with is complement $\tilde{W}'$ --- $W = \begin{bmatrix}
\tilde{W}, &
\tilde{W}'
\end{bmatrix}^\top$ --- 
 is a regular and balanced assignment matrix.
\end{lemma}

Proofs of \cref{lemma:permutation,lemma:complement} can be found in Appendix~\ref{sec:proof_lemma_permutation} and \ref{sec:proof_lemma_complement}. 
Last, we characterize in \cref{lemma:prob} the closed form probabilities of treatment assignment when $W$ is sampled according to \cref{lemma:complement}. 
This probability can be used to efficiently calculate \cref{eq:lag_p_hv_th}.

\begin{lemma}\label{lemma:prob} 
Let $W$ be drawn from a RBSD like in \cref{lemma:complement}, with $p \steptotal \in \mathbb{N}$. For $j < p \steptotal$, the probability of observing a subsequence of 1's of length $j+1$ is given by 
\[
    P(W_{\unitidx, \stepidx-j:\stepidx}=1_{\stepidx-m:\stepidx}) 
    =
    \frac{\binom{p\steptotal}{j+1}}{\binom{\steptotal}{j+1}}.
\]
\end{lemma}
See Appendix~\ref{sec:proof_lemma_prob} for a proof.
For the case of $p=\frac{1}{2}$ and $j=1$, the probability above simplifies to $P(W_{\unitidx,\stepidx-j:\stepidx}=1_{\stepidx-j:\stepidx}) = \frac{\steptotal-2}{4(\steptotal-1)}$. 
Additionally, we see that $\lim_{\steptotal\to\inf} \frac{\steptotal-2}{4(\steptotal-1)} = \frac{1}{4} = p^{j+1}$. 
In words this means that for a \textit{long running} experiment the probability of observing $j+1$ consecutive treatments (or controls) approaches that of seeing consecutive heads when a fair coin is flipped $j+1$ times.

%% file: chapters/5_simulations.tex

\section{Empirical investigations on A/A tests and semi-synthetic A/B tests} \label{sec:simulations}

In this section, we investigate the empirical performance of our proposed design on real data from an online e-commerce service.
We consider in \cref{sec:aa} data from an A/A test -- where the treatment effect is a placebo.
We use this data to empirically show that our proposed design (RBSD, \cref{sec:RBSDs}) produces estimates with lower variance than alternative available designs. 
Lower variability in the estimator naturally translates in higher efficiency, allowing practitioners to draw stronger conclusions from their experiments with the same amount of available data.

Next, we consider semi-synthetic data in \cref{sec:simulation}, where we augment the A/A test data with simulated effects.
This framework allows us to verify that even in the presence of carryover effects RBSDs yield unbiased yet precise estimators of the average treatment effects.

\subsection{
Empirical comparison of $\hat{\tau}, \hat{\tau}_\lag$ and $\hat{\sigma}, \hat{\sigma}_\lag$ across different experimental designs
}\label{sec:aa}

On a single day of sales data from an online e-commerce service, we randomly retain $\unittotal_{\mathrm{pop}} = 10,000 $ unique \ASINs{}. 
We then collect the sales data for these \ASINs{} over $\steptotal=14$ subsequent days.

Utilizing this dataset, we conduct a simulation to produce Monte Carlo estimates of the distribution of $\hat{\tau}, \hat{\tau}_\lag$ (\cref{eq:hv_th,eq:lag_p_hv_th}) and their estimated standard errors $\hat{\sigma}, \hat{\sigma}_\lag$ across draws of assignments $W$ from different experimental designs (\ASIN{}-randomization, multi-unit regular switchback,  and RBSD). We consider $\ell=1$.
Our simulation works as follows: across 100 iterations, for each design we sample at random a valid assignment $W\in\{0,1\}^{\unittotal \times \steptotal}$.
Importantly, the potential outcomes $Y_{\unitidx, \stepidx}$ are unaffected by the random assignment in this setup: the only effect of randomization is on the treatment assignment, which influence the corresponding estimators.

\begin{figure}[ht] 
\centering
\begin{subfigure}{.5\textwidth}
    \centering
    \input{figures/tau_estimate}
  \label{fig:tau_estimate}
\end{subfigure}%
\begin{subfigure}{.5\textwidth}
    \centering
    \input{figures/tau_1_estimate}
  
\end{subfigure}%
\caption{
    Results of a Monte Carlo simulation, over 100 re-draws of the assignment matrix $W$ using sales data from $2\times 10^5$ unique \ASINs{} over 14 days. 
    We report the estimated effects for $\hat{\tau}$ (left) $\hat{\tau}_1$ (right) under the same three competing experimental designs (\ASIN{}-randomized, multi-unit regular switchback, and RBSD). 
} \label{fig:tau_1_estimate}
\end{figure}
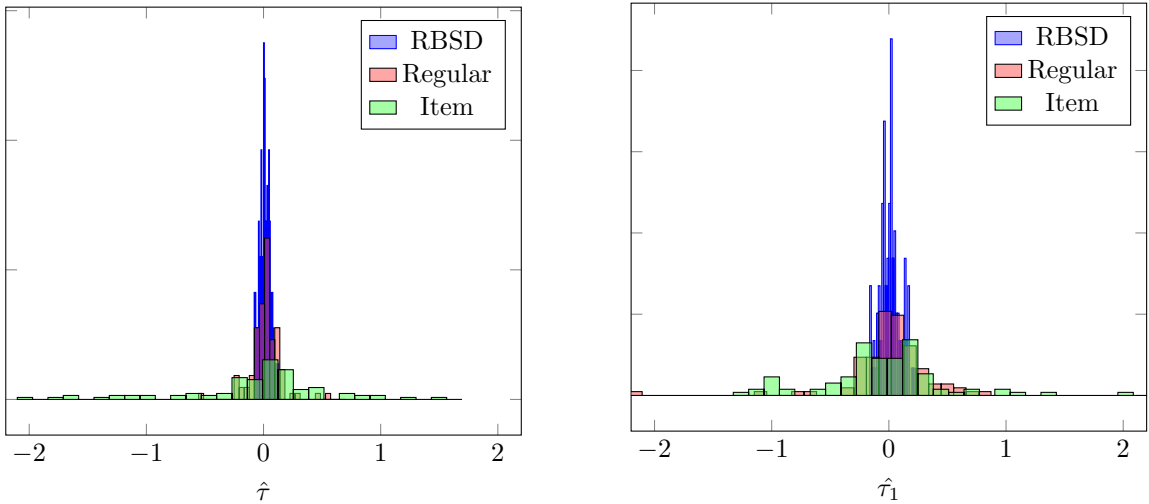

Results of the Monte Carlo simulation are reported in \cref{fig:tau_1_estimate}. 
As expected, both $\hat{\tau}$ and $\hat{\tau}_1$ are unbiased under all the experimental designs considered --- the resulting estimates are all centered around zero. 
This implicitly verifies the claims of \cref{thm:unbiased,thm:unbiased_lagp} under no effects.

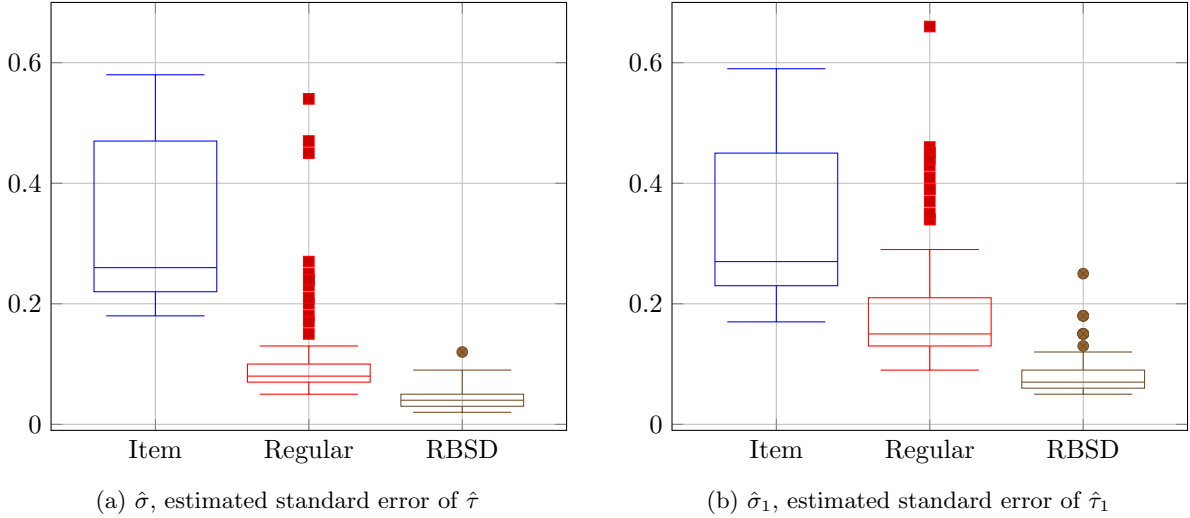
\begin{figure*}[ht]
\centering
\begin{subfigure}{.5\textwidth}
    \centering
    \input{figures/tau_std_boxplot}
  \caption{$\hat{\sigma}$, estimated standard error of $\hat{\tau}$}
  \label{fig:std_tau}
\end{subfigure}%
\begin{subfigure}{.5\textwidth}
    \centering
    \input{figures/tau_1_std_boxplot}
  \caption{$\hat{\sigma}_1$, estimated standard error of $\hat{\tau}_1$}
  \label{fig:std_tau_1}
\end{subfigure}%
\caption{Empirical distribution of 
        $\hat{\sigma}$ (left) and 
        $\hat{\sigma}_1$ (right) under the same setting of \cref{fig:tau_1_estimate}
}
\label{fig:stds}
\end{figure*}

We additionally report the empirical distribution of the estimated standard errors $\hat{\sigma}$ and $\hat{\sigma}_1$ in \cref{fig:stds}.
From the experiments, we see a clear ordering of the different experimental designs:
\ASIN{}-randomization leads to estimates that have higher variability than multi-unit switchback designs; further, RBSD produces the least variable estimates, systematically outperforming simpler multi-unit regular switchback switchback designs.
Comparing \cref{fig:std_tau} with \cref{fig:std_tau_1} we observe that $\hat{\tau}$ has a slightly smaller variance than $\hat{\tau}_1$. 
However, as we will see in the next section, the lower variance of $\hat{\tau}$ can only be exploited in cases where carryovers are not a concern.

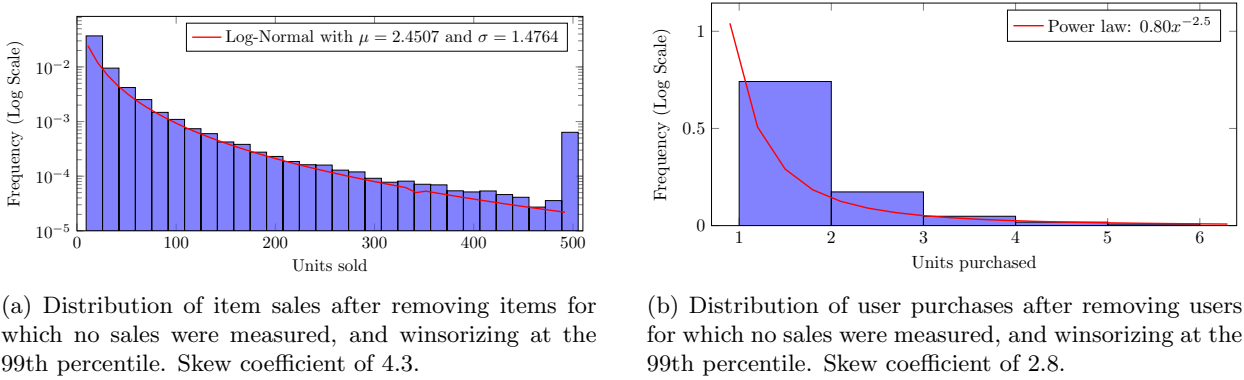
\begin{figure}
    \centering
    \begin{subfigure}[b]{0.48\textwidth}
        \centering
        \resizebox{\textwidth}{!}{\input{figures/asins_log_normal}}
        \caption{Distribution of \ASIN{} sales after removing \ASINs{} for which no sales were measured, and winsorizing at the 99th percentile. Skew coefficient of 4.3.}
        \label{fig:log_normal}
    \end{subfigure}
    \hfill
    \begin{subfigure}[b]{0.48\textwidth}
        \centering
        \resizebox{\textwidth}{!}{\input{figures/users_power_law}}
        \caption{Distribution of user purchases after removing users for which no sales were measured, and winsorizing at the 99th percentile. Skew coefficient of 2.8.}
        \label{fig:users_power_law}
    \end{subfigure}
    \caption{Distributions of ASIN sales and user purchases in the A/A data described in \cref{sec:aa}.}
    \label{fig:combined_distributions}
\end{figure}

The empirical distributions of sales aggregated at the \ASIN{} and user levels provide insight into the findings of \cref{fig:stds}. 
\Cref{fig:log_normal} provides the distribution of \ASIN{}-level sales after excluding \ASINs{} for which no purchase was made and winsorizing at the 99-th percentile. 
Similarly, \cref{fig:users_power_law} depicts user-level sales, employing the same data preprocessing steps.
Both distributions exhibit pronounced positive skewness, as quantified by the Fisher-Pearson coefficient of skewness (4.3 for \ASINs{} and 2.8 for users). 
This indicates a right-skewed distribution where a small proportion of \ASINs{} and users account for a disproportionately large share of total sales volume. 
The higher skewness observed at the \ASIN{}-level provides an explanation of why switchback randomization is more effective than plain \ASIN{}-randomization for this data.
This is in line with the previous observations of \citet{masoero2023efficient} -- by balancing out the treatment assignments over both timesteps and \ASINs{}, we reduce the impact of the highly skewed distribution on the variance of our estimates.

\subsection{Empirical validation with simulated effects} \label{sec:simulation}


Next, we test the performance of the same designs considered in \cref{sec:aa} on semi-synthetic data.
Namely, we consider the same sales data of \cref{sec:aa}, and ``augment'' it by injecting treatment effects via a linear additive model with carryover effects.
This model is standard in the causal inference literature, and has extensively been used (see, e.g., \citet{hedayat1978repeated, bojinov2023design}). 
Specifically, letting $Y^{\mathrm{A/A}}_{\unitidx, \stepidx} \in \RR$ denote the observed A/A measurement for unit $\unitidx$ at time step  $\stepidx$, and given an arbitrary assignment matrix $W \in \{0,1\}^{\unittotal, \steptotal}$, we generate potential outcomes according to the following specification: 
\begin{equation} \label{eq:lin_add_effect}
    Y_{\unitidx, \stepidx}(W_{\unitidx, 1:\stepidx}) = 
    Y^{\mathrm{A/A}}_{\unitidx, \stepidx}+ 
    \sum_{j = 0}^\stepidx \delta^{(j)} W_{\unitidx, s-j}.
\end{equation}
Under this data generating mechanism $\delta^{(0)} \in \RR$ encodes the direct treatment effect of exposure and $\delta^{(\stepidx)} \in \RR$ the $\stepidx$-th order carryover effect. 
Note that this definition of effects satisfies \cref{as:sutva} and \cref{as:potential}, but violates \cref{as:no_carryover} if $\delta^{(j)} > 0$ for any $j>0$. It also does not satisfy \cref{as:carryover} for a given carryover order $m$ if $\delta^{(j)} > 0$ for any $j>m$.
We consider a first-order carryover model, i.e.\ $\delta^{(j)}=0, \forall j >1$. 
We sample 100 assignment matrices $W \in \{0,1\}^{\unittotal_{\mathrm{pop}}, 14}$ for 4 different choices of $\delta^{(0)}$ and $\delta^{(1)}$, 2 estimators $\hat{\tau}$ and $\hat{\tau}_1$ and 3 designs (\ASIN{}-randomized, Regular multi-unit switchback and RBSD) leading to 2400 total simulations and report 
\begin{itemize}
    \item mean error (ME), to validate the unbiasedness of our estimator
    \item mean squared error (MSE), to investigate the expected distance to the true effect size
    \item false positive rate (FPR); when $\sum_j\delta^{(j)} = 0$, to show that our variance estimators of \cref{thm:unbiased,thm:unbiased_lagp} is conservative
    \item false negative rate (FNR) if $\sum_j\delta^{(j)} \ne 0$, to show the power of the design and estimator.
\end{itemize}

FNR and FPR are calculated using a two-sided one-sample z-test, testing whether the estimated impact is significantly different from zero. The test is conducted with a significance level of $\alpha = 0.05$.

\input{chapters/5_simulations_table}

\subsubsection*{Mean Error}

Looking at the mean error metric, we can see that the simplest estimator $\hat{\tau}$ from equation \cref{eq:hv_th} is biased for both the multi-unit regular switchback and RBSD designs. Specifically, the row for the ME metric where where $\delta^{(0)} = 0.0$ $\delta^{(1)} = 0.2$ shows a mean error of -0.21 for RBSD and -0.2 for Regular designs. The row with $\delta^{(0)} = 0.2$ $\delta^{(1)} = 0.2$ shows a mean error of -0.211 for RBSD and -0.199 for Regular designs. All these biases are close to the carryover effect, indicating they are biased to be completely blind to carryover effects. The \ASIN{}-randomized design, however, doesn't suffer from such a strong bias. This is because only the first of the 14 days lacks the carryover effect (consider: On the first day of an experiment, the units were not treated before, and so there can be no 1st order carryover effect). The estimate therefore is theoretically biased for 1/14th of the carryover effect. In practice this bias is often too small to be considered for long-running experiments.

The results of $\hat{\tau}_1$ show no such obvious bias, which empirically strengthen our belief that the claim of \cref{thm:unbiased_lagp} holds in these simulated conditions.

\subsubsection*{Mean Squared Error}
The MSE in absence of carryover effects (rows where $\delta^{(1)}=0$), confirms the results from \cref{sec:aa}: both RBSD and multi-unit regular switchback designs yield estimators with far lower variance than conventional \ASIN{}-randomization. 
We saw in the ME results that when $\delta^{(1)}>0$, RBSD and regular designs are biased when using estimator $\hat{\tau}$, which in turn also increases the MSE. For $\hat{\tau}_1$, no such bias was observed and indeed the MSE of $\hat{\tau}_1$ for RBSD and regular designs has no noticeable increase when $\delta^{(1)}>0$.

\subsubsection*{False Positive Rate}

We see that for all designs, the FPR does not exceed 0.05. Since our type-one error rate in the test was set at $\alpha=0.05$, these results highlight that the variance calculated per \cref{thm:unbiased} or \cref{thm:unbiased_lagp} is conservative.

\subsubsection*{False Negative Rate}

The high variance of estimators produced via the \ASIN{}-randomized design makes them underpowered to detect effects of the magnitude simulated here. 
We specifically call out the low FNR of $\hat{\tau}$ when using RBSDs and multi-unit regular switchback designs even when $\delta^{(0)}=0.2$ and $\delta^{(1)}=0.2$. 
Under these parameters configuration, $\hat{\tau}$ is biased under both designs, and still the FNR is $0.04$ for RBSD and $0.34$ for multi-unit regular switchback designs. 
We note that this is purely due to the direct effect also being present, since the FNR respectively increases to $0.98$ and $0.97$ when the direct effect is removed.
Out of all results, we see that RBSD designs consistently have the lowest FNR, with the exception of estimator $\hat{\tau}$ when $\delta^{(0)}=0.0$ and $\delta^{(1)}=0.2$ due to the bias in that estimator for RBSD designs.

Our simulation verifies that the simple estimator $\hat{\tau}$ is biased for multi-unit regular switchback and RBSD designs when carryover effects are present. 
\ASIN{}-randomized designs do not suffer from a strong bias in this case, but still have low precision (looking at the MSE). 
The estimator $\hat{\tau}_1$ is instead unbiased, in accordance with  \cref{thm:unbiased_lagp}.
Additionally, when no effects are present, all our designs keep the false positive rate in check, staying under the desired threshold $0.05$, which suggests the proposed variance calculations are on point for our sample of $10,000$ \ASINs{}.

Overall, these results suggest that RBSDs can yield estimators that are unbiased in the presence of carryover effects while still being able to detect treatment effects reliably and low MSE. 
The \ASIN{}-randomized design is robust, but lacks the statistical power of the designs also randomizing  over time. 
The simple estimator $\hat{\tau}$ is more precise than $\hat{\tau}_1$ (as per MSE and FNR) when no carryover effects are present, but can suffer from high bias in those settings.

%% file: figures/tau_estimate.tex
\begin{tikzpicture}
\begin{axis}[
    xlabel={$\hat{\tau}$},
    xmin=-2.2,
    xmax=2.2,
    xtick={-2,-1,0,1,2},
    yticklabels={},  
    legend pos=north east,
    legend style={
        /tikz/every even column/.append style={column sep=0.5cm},
        cells={anchor=center}, 
    },
    legend image code/.code={
        \draw[#1] (-5pt,-2.5pt) rectangle (5pt,2.5pt);
    },
    legend entries={RBSD,Regular,Item},
    every legend image/.style={
        draw=black,
        fill opacity=0.5
    }
]

\addplot[ybar interval,
    bar width=0.8,
    fill=blue!70,
    draw=blue,
    fill opacity=0.5] 
    coordinates {
    (-0.08013675349652444, 4.128496567597066)
(-0.07287018574993645, 4.128496567597066)
(-0.06560361800334849, 2.7523310450647114)
(-0.05833705025676051, 0.0)
(-0.05107048251017254, 2.7523310450647114)
(-0.04380391476358456, 6.880827612661777)
(-0.03653734701699659, 4.128496567597066)
(-0.029270779270408616, 5.504662090129423)
(-0.022004211523820642, 9.633158657726488)
(-0.014737643777232669, 5.504662090129423)
(-0.007471076030644695, 5.504662090129423)
(-0.00020450828405672167, 13.761655225323555)
(0.007062059462531252, 12.3854897027912)
(0.014328627209119225, 6.880827612661777)
(0.0215951949557072, 5.504662090129423)
(0.028861762702295173, 8.256993135194133)
(0.036128330448883146, 4.128496567597066)
(0.04339489819547112, 9.633158657726488)
(0.0506614659420591, 6.880827612661765)
(0.057928033688647074, 2.7523310450647163)
(0.06519460143523503, 4.128496567597074)
(0.072461169181823, 4.128496567597058)
(0.079727736928411, 2.7523310450647056)
(0.08699430467499897, 0.0)
(0.09426087242158693, 1.3761655225323581)
(0.1015274401681749, 0.0)
(0.10879400791476289, 1.3761655225323528)
(0.11606057566135086, 0.0)
(0.12332714340793882, 1.3761655225323581)
(0.13059371115452678, 1.3761655225323528)
};
\addlegendentry{RBSD}

\addplot[ybar interval,
    bar width=0.8,
    fill=red!70,
    draw=black,
    fill opacity=0.5] 
    coordinates {
    (-0.5548360925801159, 0.23059545438100837)
(-0.5114701033118401, 0.0)
(-0.4681041140435642, 0.0)
(-0.4247381247752884, 0.0)
(-0.38137213550701254, 0.0)
(-0.3380061462387367, 0.0)
(-0.29464015697046086, 0.0)
(-0.251274167702185, 0.9223818175240335)
(-0.20790817843390919, 0.46119090876201674)
(-0.16454218916563335, 0.46119090876201674)
(-0.1211761998973575, 0.9223818175240335)
(-0.07781021062908167, 2.7671454525721004)
(-0.03444422136080583, 3.689527270096134)
(0.008921767907470013, 6.226077268287225)
(0.05228775717574585, 2.305954543810084)
(0.09565374644402169, 2.7671454525721004)
(0.13901973571229753, 1.152977271905042)
(0.18238572498057337, 0.0)
(0.2257517142488492, 0.23059545438100837)
(0.26911770351712505, 0.23059545438100837)
(0.3124836927854009, 0.0)
(0.35584968205367673, 0.0)
(0.39921567132195257, 0.0)
(0.4425816605902284, 0.23059545438100837)
(0.48594764985850425, 0.0)
(0.5293136391267801, 0.23059545438100837)
(0.5726796283950559, 0.0)
(0.6160456176633318, 0.0)
(0.6594116069316076, 0.0)
(0.7027775961998834, 0.23059545438100898)
};
\addlegendentry{Regular}

\addplot[ybar interval,
    bar width=0.8,
    fill=green!70,
    draw=black,
    fill opacity=0.5] 
    coordinates {
    (-2.101639245272609, 0.07640251124074726)
(-1.9707534950845083, 0.0)
(-1.8398677448964076, 0.07640251124074726)
(-1.708981994708307, 0.15280502248149427)
(-1.578096244520206, 0.0)
(-1.4472104943321054, 0.07640251124074726)
(-1.3163247441440047, 0.15280502248149452)
(-1.1854389939559038, 0.15280502248149427)
(-1.0545532437678031, 0.15280502248149452)
(-0.9236674935797023, 0.0)
(-0.7927817433916016, 0.15280502248149452)
(-0.6618959932035009, 0.22920753372224179)
(-0.5310102430154001, 0.15280502248149427)
(-0.40012449282729934, 0.22920753372224179)
(-0.26923874263919856, 0.8404276236482185)
(-0.13835299245109778, 0.7640251124074727)
(-0.007467242262997109, 1.5280502248149455)
(0.12341850792510356, 1.146037668611209)
(0.25430425811320445, 0.38201255620373503)
(0.38519000830130534, 0.45841506744448357)
(0.516075758489406, 0.0)
(0.6469615086775067, 0.22920753372224179)
(0.7778472588656073, 0.15280502248149452)
(0.908733009053708, 0.15280502248149452)
(1.039618759241809, 0.0)
(1.1705045094299098, 0.07640251124074726)
(1.3013902596180105, 0.0)
(1.4322760098061111, 0.07640251124074726)
(1.5631617599942118, 0.0)
(1.6940475101823127, 0.229207533722241)
};
\addlegendentry{\AASIN{}}

\end{axis}
\end{tikzpicture}

%% file: figures/tau_1_estimate.tex
\begin{tikzpicture}
\begin{axis}[
    xlabel={$\hat{\tau_1}$},
    xmin=-2.2,
    xmax=2.2,
    xtick={-2,-1,0,1,2},
    yticklabels={},  
    legend pos=north east,
    legend style={
        /tikz/every even column/.append style={column sep=0.5cm}
    },
    legend image code/.code={
        \draw[#1] (-5pt,-2.5pt) rectangle (5pt,2.5pt);
    },
    legend entries={RBSD,Regular,\ASIN{}},
    every legend image/.style={
        draw=black,
        fill opacity=0.5
    }
]

\addplot[ybar interval,
    bar width=0.8,
    fill=blue!70,
    draw=blue,
    fill opacity=0.5] 
    coordinates {
    (-0.16435380578100217, 2.7012498130484137)
(-0.14954584547804184, 0.6753124532621034)
(-0.13473788517508145, 1.3506249065242069)
(-0.1199299248721211, 0.6753124532621034)
(-0.10512196456916074, 2.02593735978631)
(-0.09031400426620038, 2.7012498130484137)
(-0.07550604396324002, 1.3506249065242069)
(-0.060698083660279664, 4.727187172834723)
(-0.045890123357319304, 6.753124532621033)
(-0.031082163054358944, 2.7012498130484137)
(-0.016274202751398584, 3.3765622663105166)
(-0.0014662424484382236, 4.727187172834723)
(0.013341717854522137, 8.779061892407343)
(0.028149678157482497, 3.3765622663105166)
(0.04295763846044286, 4.05187471957262)
(0.05776559876340322, 2.02593735978631)
(0.07257355906636359, 2.0259373597863064)
(0.08738151936932395, 1.3506249065242093)
(0.10218947967228428, 1.3506249065242093)
(0.11699743997524464, 1.3506249065242042)
(0.13180540027820503, 3.3765622663105104)
(0.1466133605811654, 1.3506249065242093)
(0.16142132088412572, 2.7012498130484186)
(0.17622928118708608, 0.0)
(0.19103724149004647, 0.6753124532621021)
(0.20584520179300683, 0.6753124532621047)
(0.22065316209596716, 0.0)
(0.23546112239892752, 0.0)
(0.2502690827018879, 0.0)
(0.26507704300484825, 0.6753124532621047)
};
\addlegendentry{RBSD}

\addplot[ybar interval,
    bar width=0.8,
    fill=red!70,
    draw=black,
    fill opacity=0.5] 
    coordinates {
    (-2.2120961795124297, 0.09404340407405344)
(-2.1057622999760657, 0.0)
(-1.9994284204397013, 0.0)
(-1.8930945409033373, 0.0)
(-1.7867606613669733, 0.0)
(-1.680426781830609, 0.0)
(-1.574092902294245, 0.0)
(-1.467759022757881, 0.0)
(-1.361425143221517, 0.0)
(-1.2550912636851526, 0.0)
(-1.1487573841487886, 0.09404340407405344)
(-1.0424235046124246, 0.0)
(-0.9360896250760604, 0.0)
(-0.8297557455396963, 0.09404340407405325)
(-0.7234218660033322, 0.09404340407405344)
(-0.6170879864669682, 0.0)
(-0.510754106930604, 0.0)
(-0.4044202273942399, 0.18808680814810688)
(-0.2980863478578759, 0.9404340407405344)
(-0.19175246832151172, 0.9404340407405306)
(-0.08541858878514752, 2.068954889629176)
(0.020915290751216453, 1.9749114855551224)
(0.12724917028758043, 1.2225642529626948)
(0.23358304982394462, 0.6583038285183714)
(0.3399169293603088, 0.28213021222216034)
(0.4462508088966728, 0.28213021222216034)
(0.552584688433037, 0.1880868081481061)
(0.6589185679694012, 0.09404340407405344)
(0.7652524475057652, 0.09404340407405344)
(0.8715863270421292, 0.09404340407405325)
};
\addlegendentry{Regular}

\addplot[ybar interval,
    bar width=0.8,
    fill=green!70,
    draw=black,
    fill opacity=0.5] 
    coordinates {
    (-1.3280346767694882, 0.0761579755900473)
(-1.196728665514288, 0.15231595118009433)
(-1.0654226542590877, 0.45694785354028383)
(-0.9341166430038876, 0.15231595118009447)
(-0.8028106317486876, 0.0)
(-0.6715046204934875, 0.1523159511800946)
(-0.5401986092382873, 0.30463190236018894)
(-0.40889259798308725, 0.4569478535402834)
(-0.27758658672788716, 1.294685585030804)
(-0.14628057547268702, 0.9138957070805661)
(-0.014974564217486885, 0.9138957070805677)
(0.11633144703771314, 1.3708435606208513)
(0.24763745829291328, 0.5331058291303302)
(0.3789434695481134, 0.1523159511800946)
(0.5102494808033136, 0.07615797559004717)
(0.6415554920585137, 0.1523159511800946)
(0.7728615033137138, 0.0)
(0.9041675145689139, 0.15231595118009486)
(1.035473525824114, 0.07615797559004717)
(1.1667795370793141, 0.0)
(1.2980855483345142, 0.07615797559004743)
(1.4293915595897142, 0.0)
(1.5606975708449145, 0.0)
(1.6920035821001145, 0.0)
(1.8233095933553145, 0.0)
(1.9546156046105148, 0.07615797559004717)
(2.085921615865715, 0.0)
(2.217227627120915, 0.0)
(2.3485336383761153, 0.0)
(2.479839649631315, 0.07615797559004743)
};
\addlegendentry{\AASIN{}}

\end{axis}
\end{tikzpicture}

%% file: figures/tau_std_boxplot.tex
\begin{tikzpicture}
  \begin{axis}
    [
    ymin=-0.01,
    ymax=0.7,
    xtick={1,2,3},
    xticklabels={\AASIN{}, Regular, RBSD},
    boxplot/draw direction=y,  
    grid=major,               
    ]
    \addplot+[
    boxplot prepared={
    median=0.26,
    upper quartile=0.47,
    lower quartile=0.22,
    upper whisker=0.58,
    lower whisker=0.18
    },
    ] coordinates { (3,1.19) (3,1.44) (3,1.47) (3,1.46) (3,1.44) (3,1.48) (3,1.55) (3,1.49) (3,1.43) (3,1.04) (3,1.07) (3,1.06) (3,1.00) (3,1.21) (3,1.22) (3,1.20) (3,1.48) (3,1.44) (3,1.19) (3,1.22) (3,1.20) (3,1.46)};
    \addplot+[
    boxplot prepared={ 
        median=0.08,
        upper quartile=0.10,
        lower quartile=0.07,
        upper whisker=0.13,
        lower whisker=0.05
    },
    ] coordinates {(2,0.24) (2,0.47) (2,0.25) (2,0.20) (2,0.15) (2,0.23) (2,0.27) (2,0.54) (2,0.45) (2,0.18) (2,0.17) (2,0.71) (2,0.24) (2,0.21)}; 
    \addplot+[
    boxplot prepared={
        median=0.04,
        upper quartile=0.05,
        lower quartile=0.03,
        upper whisker=0.09,
        lower whisker=0.02
    },
    ] coordinates {(1,0.12)}; 
  \end{axis}
\end{tikzpicture}

%% file: figures/tau_1_std_boxplot.tex
\begin{tikzpicture}
  \begin{axis}
    [
    ymin=-0.01,
    ymax=0.7,
    xtick={1,2,3},
    xticklabels={\AASIN{}, Regular, RBSD},
    boxplot/draw direction=y,  
    grid=major,               
    ]
    \addplot+[
    boxplot prepared={
        median=0.27,
        upper quartile=0.45,
        lower quartile=0.23,
        upper whisker=0.59,
        lower whisker=0.17,
      draw position=1
    },
    ] coordinates {(3,1.15) (3,0.97) (3,0.99) (3,1.16) (3,0.97) (3,0.95) (3,0.98) (3,1.55) (3,0.96) (3,1.02) (3,0.96) (3,1.38) (3,1.49) (3,1.40) (3,0.96) (3,1.22)};
    \addplot+[
    boxplot prepared={ 
    median=0.15,
    upper quartile=0.21,
    lower quartile=0.13,
    upper whisker=0.29,
    lower whisker=0.09,
      draw position=2
    },
    ] coordinates {(2,0.66) (2,0.46) (2,0.45) (2,0.34) (2,0.37) (2,0.39) (2,1.28) (2,0.34) (2,0.43) (2,0.73) (2,0.78) (2,0.45) (2,0.44) (2,2.09) (2,0.39) (2,0.41) (2,0.88) (2,0.35)}; 
    \addplot+[
    boxplot prepared={
    median=0.07,
    upper quartile=0.09,
    lower quartile=0.06,
    upper whisker=0.12,
    lower whisker=0.05,
      draw position=3
    },
    ] coordinates {(1,0.15) (1,0.13) (1,0.15) (1,0.18) (1,0.15) (1,0.18) (1,0.25) (1,0.15) (1,0.15) (1,0.18)}; 
  \end{axis}
\end{tikzpicture}

%% file: figures/asins_log_normal.tex
\begin{tikzpicture}
\begin{axis}[
    width=12cm,
    height=6cm,
    xlabel={Units sold},
    ylabel={Frequency (Log Scale)},
    ymode=log,
    log origin=infty,  
    legend pos=north east,
    ymin=0.00001,  
    xmin=0,        
    xmax=510,      
    xtick={0,100,200,300,400,500}, 
    scaled x ticks=false,  
    /pgf/number format/.cd,
    fixed,
    precision=0
]

\addplot[ybar interval,fill=blue!70,draw=black,fill opacity=0.7, forget plot] coordinates {
(9.266666666666667, 0.03695608986409348)
(25.800000000000004, 0.0095017688971523)
(42.33333333333334, 0.004193218553935156)
(58.866666666666674, 0.002523080695112902)
(75.4, 0.0014728099268723658)
(91.93333333333334, 0.0010945980573017428)
(108.46666666666668, 0.0007396222666744963)
(125.00000000000001, 0.0006005632711518384)
(141.53333333333336, 0.0004228966367694196)
(158.06666666666666, 0.0003821441290841163)
(174.60000000000002, 0.00027525816594459266)
(191.13333333333335, 0.00022985844247061396)
(207.66666666666669, 0.00018588863154699763)
(224.20000000000005, 0.00016193759632844216)
(240.73333333333335, 0.00015979272750289987)
(257.2666666666667, 0.0001290496076701272)
(273.80000000000007, 0.00011975517609277711)
(290.33333333333337, 9.151440322313745e-05)
(306.8666666666667, 7.757275585711234e-05)
(323.40000000000003, 8.114753723301642e-05)
(339.9333333333334, 7.113814938048552e-05)
(356.4666666666667, 6.935075869253385e-05)
(373.0, 5.3979198776147295e-05)
(389.5333333333334, 5.1476851813014635e-05)
(406.0666666666667, 5.36217206385571e-05)
(422.6, 4.611467974915895e-05)
(439.1333333333334, 4.110998582289378e-05)
(455.66666666666674, 2.716833845686884e-05)
(472.20000000000005, 3.574781375903807e-05)
(488.73333333333335, 0.0006355961286356968)
(505.26666666666665, 0.0006355961286356968)  
};

\addplot[red,thick] coordinates {

(11.02020202020202, 0.02450474910886462)
(21.04040404040404, 0.011838476914847441)
(31.060606060606062, 0.0069630764469475834)
(41.08080808080808, 0.0045572574132529305)
(51.101010101010104, 0.0031927693551279442)
(61, 0.002345877146928077)
(71.14141414141415, 0.0017856944100263436)
(81.16161616161617, 0.001397099548371965)
(91.18181818181819, 0.0011173586112897978)
(101.20202020202021, 0.0009098989794892217)
(111.22222222222223, 0.0007522351188274464)
(121.24242424242425, 0.0006299338771584722)
(131.26262626262627, 0.000533394316827026)
(141.2828282828283, 0.0004560349319433074)
(151.3030303030303, 0.0003932265352042103)
(161.32323232323233, 0.0003416398332498638)
(171.34343434343435, 0.0002988333286977264)
(181.36363636363637, 0.0002629855396671557)
(191.3838383838384, 0.00023271661048109284)
(201.40404040404042, 0.00020696682291839217)
(211.42424242424244, 0.00018491221408031017)
(221.44444444444446, 0.00016590492195484504)
(231.46464646464648, 0.00014943033333031562)
(241.4848484848485, 0.00013507585167049624)
(251.50505050505052, 0.000122507830702563)
(261.52525252525254, 0.00011145433086710355)
(271.54545454545456, 0.00010169208406452779)
(281.5656565656566, 9.30365376283194e-05)
(291.5858585858586, 8.533417724624586e-05)
(301.6060606060606, 7.845655447931943e-05)
(311.62626262626264, 7.229560187964499e-05)
(321.64646464646466, 6.67599296762597e-05)
(331.6666666666667, 6.177187717121253e-05)
(341.6868686868687, 5e-05)
(351.7070707070707, 5.318290876457461e-05)
(361.72727272727275, 4.9476230545748086e-05)
(371.74747474747477, 4.610283673775679e-05)
(381.7676767676768, 4.302606112943961e-05)
(391.7878787878788, 4.021399420418818e-05)
(401.80808080808083, 3.763877477438441e-05)
(411.82828282828285, 3.5276000278092517e-05)
(421.8484848484849, 3.310423372765006e-05)
(431.8686868686869, 3.11045897641315e-05)
(441.8888888888889, 2.9260385755859853e-05)
(451.90909090909093, 2.7556846614193164e-05)
(461.92929292929296, 2.598085415866361e-05)
(471.949494949495, 2.4520733576585523e-05)
(481.969696969697, 2.3166070888385356e-05)
(491.989898989899, 2.1907556424643573e-05)

};
\addlegendentry{Log-Normal with $\mu=2.4507$ and $\sigma=1.4764$}
\end{axis}
\end{tikzpicture}

%% file: figures/users_power_law.tex
\begin{tikzpicture}
\begin{axis}[
    width=12cm,
    height=6cm,
    xlabel={Units purchased},
    ylabel={Frequency (Log Scale)},
    legend pos=north east,
    legend style={
        /tikz/every even column/.append style={column sep=0.5cm}
    },
    ymin=0,  
    xmin=0.7,        
    xmax=6.4,      
    xtick={1,2,3,4,5,6}, 
    scaled x ticks=false, 
    /pgf/number format/.cd,
    fixed,
    precision=1
]

\addplot[ybar interval,bar width=0.8,fill=blue!70,draw=black,fill opacity=0.7, forget plot] coordinates {
(1, 0.741567)
(2, 0.173402)
(3, 0.048140)
(4, 0.017885)
(5, 0.008096)
(6, 0.010909)
};
\addplot[red,thick] coordinates {
(0.8999999999999999, 1.04107906507601)
(1.2, 0.5071505162084872)
(1.5, 0.29030989544096925)
(1.7999999999999998, 0.18403851666664942)
(2.1, 0.1251819608784316)
(2.4, 0.08965239227331985)
(2.6999999999999997, 0.06678525316325759)
(3.0, 0.05132002392796673)
(3.3, 0.040439440408681755)
(3.5999999999999996, 0.03253372078362531)
(3.9, 0.026633513927249622)
(4.2, 0.022129253354842022)
(4.5, 0.01862338847569508)
(4.8, 0.015848453631515224)
(5.1, 0.013619606082892664)
(5.3999999999999995, 0.01180607634874994)
(5.7, 0.010313423411306753)
(6.0, 0.009072184232530289)
(6.3, 0.008030426534538634)
};
\addlegendentry{Power law: $0.80 x^ {-2.5}$}

\end{axis}
\end{tikzpicture}

%% file: chapters/5_simulations_table.tex
\begin{table}[htbp]
  \centering
  \begin{tabular}{l|ll|ccc|ccc}
    & \multicolumn{5}{c|}{$\hat{\tau}$} & \multicolumn{3}{c}{$\hat{\tau}_1$} \\
    \hline
    & \multirow{2}{*}{\textbf{$\delta^{(0)}$}} & 
      \multirow{2}{*}{\textbf{$\delta^{(1)}$}} & 
      \multicolumn{3}{c|}{\textbf{Experimental Design}} &
      \multicolumn{3}{c}{\textbf{Experimental Design}} \\
    \cline{4-9}
    \textbf{Metric} & & & \textbf{\AASIN{}} & \textbf{RBSD} & \textbf{Regular} &
    \textbf{\AASIN{}} & \textbf{RBSD} & \textbf{Regular} \\
    \hline
    \multirow{4}{*}{ME}
    & 0.0 & 0.0 & -0.079 & 0.011 & 0.015 & -0.038 & 0.016 & 0.026 \\
    & 0.0 & 0.2 & -0.009 & -0.21 & -0.2 & 0.03 & 0.013 & 0.017 \\
    & 0.2 & 0.0 & -0.03 & -0.011 & -0.004 & -0.013 & 0.014 & -0.034 \\
    & 0.2 & 0.2 & -0.011 & -0.211 & -0.199 & 0.064 & 0.014 & -0.021 \\
    \hline
    \multirow{4}{*}{MSE}
    & 0.0 & 0.0 & 0.388 & 0.003 & 0.016 & 0.318 & 0.011 & 0.062 \\
    & 0.0 & 0.2 & 0.396 & 0.046 & 0.073 & 0.289 & 0.01 & 0.065 \\
    & 0.2 & 0.0 & 0.286 & 0.002 & 0.03 & 0.339 & 0.008 & 0.068 \\
    & 0.2 & 0.2 & 0.342 & 0.047 & 0.074 & 0.319 & 0.012 & 0.075 \\
    \hline
    FPR & 0.0 & 0.0 & 0.03 & 0.02 & 0.03 & 0.02 & 0.02 & 0.01 \\
    \cline{1-9}
    \multirow{3}{*}{FNR}
    & 0.0 & 0.2 & 0.83 & 0.98 & 0.97 & 0.95 & 0.25 & 0.72 \\
    & 0.2 & 0.0 & 0.94 & 0.1 & 0.34 & 0.92 & 0.24 & 0.72 \\
    & 0.2 & 0.2 & 0.65 & 0.04 & 0.34 & 0.67 & 0.07 & 0.29 \\
    \hline
  \end{tabular}
  \caption{Comparison of \ASIN{}-randomized, multi-unit regular switchback and RBSD across different parameter configurations for simulated treatment effects $\delta^{(0)}, \delta^{(1)}$ on performance metrics of interest.
}
  \label{tab:simulated}
\end{table}

%% file: chapters/6_experiments.tex
\section{Online Experimentation} \label{sec:exp_real}

In this section, we consider the results of multiple A/B test run on an online e-commerce service.
Since the effects here considered are not simulated, no ground truth measurement is available to establish the quality of our findings. 
For that reason, we run each experiment using alternative experimental designs and adopt \textit{method triangulation} as proposed by \citet{carter2014use}. Specifically, both experiments test the same change and thus we expect the same item‐level average treatment effect (as formalized in \cref{eq:tau} and \cref{eq:lag_p_tau}), but the experiments differ in their randomization unit or randomization method. We compare estimates from both designs to strengthen our belief that our approach correctly estimates the average treatment effect.

\subsection{
    Online experiment 1: Human verification of automatically translated content
}
\label{sec:MTPE_experiment}

The first intervention tests the impact of human \textit{post-editing} of machine translated content. Post-editing is a setup where machine translation is applied in bulk to monolingual content, and human experts are asked to verify and edit the machine translations. See e.g. \cite{green2013efficacy} for a detailed review of this process. 
Post-editing is expected to improve the quality of the translations, but comes at a cost.
To measure its value, we select $40,581$ \ASINs{} at random and use automated machine translation to translate the \ASIN{} content. 
This translated content is sent to highly skilled human translators for a review and edited where necessary. 
The translated content of \ASINs{} has predominantly a front-end impact: users can change the {language of preference} of the e-commerce service they interact with, and all content will be shown in that language. 
There are also no legal or ethical concerns in experimenting with translations, as such we can run a \textit{user-randomized} experiment for triangulation. We experiment for eight weeks, during which each user is randomly assigned into a control or treatment group according to a fair coin. 
When a treated user interacts with one of the $40,581$ selected \ASINs{}, they see a human curated version of its content. 
Conversely, users in control see machine-translated content across all \ASINs{}.

Secondly, we run a regular balanced switchback design on the same $40,581$ \ASINs{} over fourteen subsequent days, not overlapping with the user-randomized experiment.
To estimate the causal effect, we use $\hat{\tau}_1$ (\cref{eq:lag_p_hv_th}), and assume the carryover effects are limited to 1 day. 
For this setup that implies $\unittotal =40,581$, $\steptotal=14$ and $\lag=1$. 

The analysis methods differ between user-randomized and switchback experiments due to their designs. For user-randomized experiments, we employ a two-sample z-test comparing the means of the treatment and control groups, since we have two distinct populations (treated and untreated users) with separate means to compare. For the RBSD experiment we use a one-sample z-test around 0 since, the estimators $\hat{\tau}$ and $\hat{\tau}_1$ estimate the treatment effect within each unit over time, resulting in a single population of difference estimates. The null hypothesis here is that the average difference (treatment effect) is zero. Both tests use z-statistics due to large sample sizes.
Results of both methods are shown in \cref{tab:mtpe}.

\begin{table}[htbp]
    \centering    
    \caption{Human curation of machine-translated content, significant results in bold}
    \label{tab:mtpe}

    \begin{tabular}{lc|c}
        \toprule
        \textbf{Metric} & \textbf{User-randomized} & \textbf{RBSD} \\
        \midrule
        Dollar uplift & 0.70\% [-1.34\%, 2.74\%] & \textbf{2.49\% [0.80\%, 4.19\%]} \\
        \cmidrule(lr){2-3}
        Conversion uplift & \textbf{1.75\% [0.35\%, 3.14\%]} & \textbf{2.51\% [0.96\%, 4.05\%]} \\
        \bottomrule
    \end{tabular}
\end{table}

A first observation is that a 14 day switchback experiment gives us confidence intervals that are similar in width and overlap with the 8 week user-randomized experiment.
While it is not generally true that \ASIN{}-randomization allows for quicker decision making, one advantage over user-randomization is that it allows to achieve higher sample sizes in shorter horizons.
Indeed, in \ASIN{}-randomization all experimental units are available from the first day of the experiment.
Conversely, in a user-randomized experiment the number of units (users triggering in the experiment) grows over time.
In particular, this implies that in certain settings longer horizon experiments might be needed to achieve desirable levels of statistical significance.
The empirical results of our experiment reveal that both user-randomized and  RBSDs  estimate a positive impact of post-editing machine translated content.

\subsection{Impact of improved machine learning model for \ASIN{} display titles in an online service}
\label{sec:title_experiment}

The second intervention tests the impact of using a machine learning model to align \ASIN{} display titles with detailed attributes in our e-commerce service. Since \ASIN{} content is manually specified by sellers it is not uncommon for e.g. a brand to be part of the title, but that same brand could be missing in the detailed attributes. 
The model performance was assessed using two different experimental designs: \ASIN{}-randomization and regular balanced switchback. In this experiment, the experimenters were particularly concerned about interference across units, and as such we follow \cref{remark:sutva} in the setup. We initially take the total population of 1,300,319 \ASINs{}, and cluster it into 99{,}443 \textit{\ASIN{}-families}. 
Each of these \textit{families} represents similar products (different sizes of the same shirt, different colors of the same smartphone, etc.). As such, these experiments are \textit{clustered} \ASIN{}-randomized and clustered regular balanced switchback.

We randomly select 17{,}000 families for \textit{regular balanced switchback} randomization for 14 timesteps spread out over 14 days. 
For all \ASINs{} in the remaining 82{,}443 families we run a four-week \ASIN{}-randomized experiment. 

To analyze the results of the \ASIN{}-randomized experiment, we use a difference-in-difference approach where we follow both the treated and untreated the families for 4 weeks before treatment. 
We use a two-sample z-test comparing the means of each group of families. 
The switchback results are again analyzed with the $\hat{\tau}_1$ estimator of \cref{eq:lag_p_hv_th} with $\unittotal=17{,}000$, $\steptotal=14$ and $\lag=1$. 
The estimate was tested using a one-sample z-test around 0.
Results are shown in \cref{tab:conventional}.

\begin{table}
    \centering
\caption{Results of online experiments. Results with $p < 0.05$ in \textbf{bold}}
\label{tab:conventional}
\begin{tabular}{ l|c c}
  & \AASIN{}-randomized & RBSD  \\ \hline
 Dollar uplift  & -2.60\% (-6.44\%, 1.14\%) & \textbf{-2.60\% (-4.43\%, -0.77\%)}  \\
 Conversion uplift  & \textbf{-4.05\% (-7.63\%, -0.46\%)}& \textbf{-2.59\% (-4.14\%, -1.04\%)} \\
\end{tabular}
\end{table}

Similar to the results of \cref{sec:MTPE_experiment}, both experimental designs yield compatible estimates of the impact of treatment.
Unlike the case of the human-curated machine translation, in this case both experimental designs randomize treatment assignment at the same unit (i.e.\ both methods treated and analyzed \ASINs{}, whereas in the case considered in \cref{sec:MTPE_experiment} we compared a user-randomized experiment to an \ASIN{}-randomized experiment). 
This allows us to highlight the predicted benefits of the \textit{regular balanced switchback} design: even though the simple \ASIN{}-randomized experiment used five times more \ASINs{}, and ran for twice as long, it yields much larger confidence intervals in the final results. 

%% file: chapters/7_conclusions.tex
\section{Conclusion} \label{sec:conclusions}

In this paper we introduced a novel experimental design --- RBSDs.
First, we established the theoretical foundations for switchback experiments across multiple units while accounting for carryover effects. 
We do so by combining research on multiple randomized designs and the prior work on switchback. 
Secondly, we show the superior performance of our design compared to simple alternatives using simulations on e-commerce data.
On the data analyzed, our design yields estimators achieving the lowest standard error against comparable designs. 
Thirdly, we propose a practical implementation through a structured algorithm for treatment allocation and a closed-form probability distribution for the proposed estimators. 
Finally, we discuss two online experiments which show our simulations carry over to realistic scenarios.

Our framework is particularly valuable for scenarios where the unit of randomization is highly skewed.
In these settings, we show that time-randomized designs can be an effective way to increase power. 
Moreover, unlike competing multi-unit designs that use time-randomization, ours properly accounts for carryover effects.

Future research will explore different statistical tests to detect when carryover effects are present, including tuning the assumed upper bound. 
We also see value in exploring the conditions under which our design is provably optimal.
Another interesting direction will be to incorporate clustering within these designs, in settings where experimenters worry that \cref{eq:SUTVA} might be violated.

Overall our work provides both theoretical guarantees and practical evidence for a novel experimental design that addresses key challenges in modern online experimentation.

%% file: chapters/9_appendix.tex
\input{chapters/9_1_appendix_equivalence}
\input{chapters/9_2_appendix_proofs}

%% file: chapters/9_1_appendix_equivalence.tex
\section{Equivalence of time-bound carryover effects and interference network}
\label{apdx:bojinov_aranov}
We compare the estimator proposed by \cite{bojinov2023design} ---which is unbiased under the assumption of time-bound carryover effects and SUTVA--- with the estimator proposed in \cite{aronow2017sutva} --- Which uses a known network of interference ---. We show that they are equivalent under a carefully chosen interference network where each observation of a unit-timestep is a node in the graph with interference edges only existing between a node and the node of 1 timestep later. 
We defer the reader to the work of Aronow for detailed discussion of the below concepts, more formally we can state that for an \textit{exposure mapping} as follows:

\begin{equation}
f(\textbf{z}, t) = \begin{cases}
d_{11}^{*}: & z_t z_{t-1} = 1 \\
d_{10}^{\dagger}: & 1(1-z_{t-1}) = 1 \\
d_{01}^{\ddagger}: & (1-z_t)z_{t-1} = 1 \\
d_{00}^{\S}: & (1-z_t)(1-z_{t-1}) = 1
\end{cases}
\end{equation}
$^{*}$Direct + Carryover, $^{\dagger}$Direct Only, 
$^{\ddagger}$Carryover Only, $^{\S}$No Exposure

and as our choice of causal impact of interest is $d_{11} - d_{00}$, i.e. the incremental impact of a full roll-out. The \textit{generalized probability of exposure} for these operands is

\begin{equation}
\begin{aligned}
\pi_t(d_{11}) &= \text{Prob}( z_t =1, z_{t-1}=1), \\
\pi_t(d_{00}) &= \text{Prob}( z_t =0, z_{t-1}=0)
\end{aligned}
\end{equation}
for which we can use the Horvitz-Thompson estimator proposed by Aronow as follows
\begin{equation}
\begin{aligned}
y(d_{11})-y(d_{00}) = \sum_{t=1}^{T} & \left\{ Y_t \frac{\mathbb{I}(z_t z_{t-1})}{\text{Prob}(z_t =1, z_{t-1}=1)} - 
 \right. \\
& \left. \quad Y_t \frac{\mathbb{I}\big((1-z_t)(1-z_{t-1})\big)}{\text{Prob}(z_t =0, z_{t-1}=0)} \right\}
\end{aligned}
\end{equation}
Which reduces to the estimator proposed by Bojinov, highlighting the network approach encompasses the simpler assumption of time-bounded carryovers.

%% file: chapters/9_2_appendix_proofs.tex
\section{Proofs} \label{sec:proofs}

\subsection{Proof of~\cref{thm:unbiased}} \label{sec:proof_unbiased}

\begin{proof}\noindent
Under \Cref{as:sutva}, we can write the expectation of \Cref{eq:hv_th} as
\begin{align*}
\EE[\hat{\tau}] &= \frac{1}{\unittotal \steptotal} \sum_{\unitidx=1}^{\unittotal}  
        \\
        &  \EE \left[ \sum_{\stepidx = 1}^{\steptotal}                     
                    \left\{ 
                        \frac{W_{\unitidx, \stepidx}}{P(W_{\unitidx, \stepidx}=1)}Y_{\unitidx, \stepidx}(W_{n,1: \steptotal}) \right. \right.\\
                        & - \left. \left. \frac{(1-W_{\unitidx, \stepidx})}{P(W_{\unitidx, \stepidx}=0)}Y_{\unitidx, \stepidx}(W_{\unitidx,1:\steptotal}) 
                    \right\}
            \right],
\intertext{and further imposing \Cref{as:potential} and \ref{as:no_carryover},}
\EE[\hat{\tau}] &= \frac{1}{\unittotal \steptotal} \sum_{\unitidx=1}^{\unittotal}  \sum_{\stepidx = 1}^{\steptotal} \\
        & \EE \left[\frac{W_{\unitidx, \stepidx}}{P(W_{\unitidx, \stepidx}=1)}Y_{\unitidx, \stepidx}(W_{\unitidx, \stepidx}) \right] \\
        & - \EE \left[ \frac{(1-W_{\unitidx, \stepidx})}{P(W_{\unitidx, \stepidx}=0)}Y_{\unitidx, \stepidx}(W_{\unitidx, \stepidx}) \right]
\end{align*}
Now, by its construction,
\[
    \EE 
    \left[ 
        \frac{W_{\unitidx, \stepidx}}{P(W_{\unitidx, \stepidx}=1)}Y_{\unitidx, \stepidx}(W_{\unitidx, \stepidx}) \right]
        = Y_{\unitidx, \stepidx}(1),
\]
and similarly $\EE 
    \left[\frac{(1-W_{\unitidx, \stepidx})}{P(W_{\unitidx, \stepidx}=0)}Y_{\unitidx, \stepidx}(W_{\unitidx, \stepidx})\right] = Y_{\unitidx, \stepidx}(0)$.
Combining the two facts,
\begin{align*}
    \EE[\hat{\tau}] &=\frac{1}{\unittotal \steptotal} \sum_{\unitidx=1}^{\unittotal} \sum_{\stepidx = 1}^{\steptotal}  Y_{\unitidx, \stepidx}(W_{\unitidx, \stepidx} =1 ) - Y_{\unitidx, \stepidx}(W_{\unitidx, \stepidx} = 0 )  \nonumber
\intertext{or, equivalently}
    \EE[\hat{\tau}] &=
        \frac{1}{\unittotal \steptotal} \sum_{\unitidx=1}^{\unittotal} \sum_{\stepidx = 1}^{\steptotal}  Y^T_{\unitidx, \stepidx} - Y^C_{\unitidx, \stepidx}. \nonumber
\end{align*}
\end{proof}

\subsection{Proof of \cref{thm:unbiased_lagp}} \label{sec:proof_unbiased_lagp}

\begin{proof} \noindent
The proof follows a similar strategy to the one presented in \Cref{thm:unbiased}. \Cref{as:sutva} allows us to express potential outcomes only as a function of the individual unit's assignments.
Moreover, \Cref{as:potential} and \Cref{as:carryover} allow us to write each potential outcome $Y_{\unitidx, \stepidx}$ as a function of only the assignments $[W_{\unitidx, \stepidx - p}, \ldots, W_{\unitidx, \stepidx}]$. 
In turn, we can write 
\begin{align*}
    \EE\left[\hat{\tau}_\lag\right] &= 
    \sum_{\unitidx=1}^\unittotal 
        \sum_{\stepidx = p+1}^{\steptotal} \\
            &\EE\left[         
                \frac{
                1(W_{\unitidx,\stepidx-p:\stepidx}=\mathbf{1}^{p+1}) Y_{\unitidx, \stepidx}(W_{\unitidx,\stepidx-p:\stepidx}=\mathbf{1}^{p+1})
                }{
                \unittotal (\steptotal -p) \; P(W_{\unitidx,\stepidx-p:\stepidx}=\mathbf{1}^{p+1})
                }
             \right. \\
             & -\left. \frac{
                1(W_{\unitidx,\stepidx-p:\stepidx}=\mathbf{0}^{p+1})  Y_{\unitidx, \stepidx}(W_{\unitidx,\stepidx-p:\stepidx}=\mathbf{0}^{p+1}) 
                }{
                \unittotal (\steptotal -p) \; P(W_{\unitidx,\stepidx-p:\stepidx}=\mathbf{0}^{p+1})
                }
           \right]
\end{align*}
By its construction,
$
    \EE\left[\frac{
                1(W_{\unitidx,\stepidx-p:\stepidx}=\mathbf{1}^{p+1}) Y_{\unitidx, \stepidx}(W_{\unitidx,\stepidx-p:\stepidx}=\mathbf{1}^{p+1})
                }{
                P(W_{\unitidx,\stepidx-p:\stepidx}=\mathbf{1}^{p+1})
                }\right] 
        = Y_{\unitidx, \stepidx}(W_{\unitidx,\stepidx-p:\stepidx}=\mathbf{1}^{p+1})
$, hence
\begin{align*}
    \EE\left[\hat{\tau}_{\lag}\right] &=
        \frac{1}{\unittotal (\steptotal -p)}\sum_{\unitidx=1}^{\unittotal} 
        \sum_{\stepidx = p+1}^{\steptotal} \\
        &
        Y_{\unitidx, \stepidx}(W_{\unitidx,\stepidx-p:\stepidx}
        =
        \mathbf{1}^{p+1})  
        - Y_{\unitidx, \stepidx}(W_{\unitidx,\stepidx-p:\stepidx}
        =\mathbf{0}^{p+1})
\end{align*}
using \Cref{as:carryover}, this  simplifies to
\[
    \EE\left[\hat{\tau}_{\lag}\right] = \frac{1}{\unittotal (\steptotal -p)} \sum_{\unitidx=1}^{\unittotal} \sum_{\stepidx = p+1}^{\steptotal}  Y^T_{\unitidx, \stepidx} - Y^C_{\unitidx, \stepidx} = \tau_p.
\]
\end{proof}

\subsection{Proof of \cref{lemma:optimal}}
\label{sec:proof_lemma_optimal}

\begin{proof} \noindent
The proof proceeds in three steps. 

First, under \cref{as:sutva}, the potential outcomes for one unit are unaffected by the treatment assignments of other units. This implies that we can consider the variance of the estimator for each unit independently. 
Secondly, due to the linearity of the estimator and the fact that the treatment assignments are independent across units, we can decompose the variance of the overall estimator as follows:
\begin{align}
\begin{split}
    \text{Var}(\hat{\tau}_{\lag}(W)) & = 
        \text{Var}\left( \right.
        \frac
            {1}
            {\unittotal(\steptotal-p)} 
        \sum_{\unitidx=1}^{\unittotal}
        \sum_{\stepidx = p+1}^{\steptotal} 
            Y_{\unitidx, \stepidx}(W) 
        \\
        & 
        \left[ 
            \frac
                {1(W_{\unitidx,\stepidx-p:\stepidx}=\mathbf{1}^{p+1})}
                {P(W_{\unitidx,\stepidx-p:\stepidx}=\mathbf{1}^{p+1})}
            -
              \frac
                {1(W_{\unitidx,\stepidx-p:\stepidx}=\mathbf{0}^{p+1})}
                {P(W_{\unitidx,\stepidx-p:\stepidx}=\mathbf{0}^{p+1})}
        \right] 
        \left. \right)\\
    \text{Var}(\hat{\tau}_{\lag}(W)) & = 
        \frac
            {1}
            {\unittotal^2} 
        \sum_{\unitidx=1}^{\unittotal}
        \text{Var}\left( \right.
        \frac{1}{(\steptotal-p)}\sum_{\stepidx = p+1}^{\steptotal} 
            Y_{\unitidx, \stepidx}(W) 
        \\
        & 
        \left[ 
            \frac
                {1(W_{\unitidx,\stepidx-p:\stepidx}=\mathbf{1}^{p+1})}
                {P(W_{\unitidx,\stepidx-p:\stepidx}=\mathbf{1}^{p+1})}
            -
              \frac
                {1(W_{\unitidx,\stepidx-p:\stepidx}=\mathbf{0}^{p+1})}
                {P(W_{\unitidx,\stepidx-p:\stepidx}=\mathbf{0}^{p+1})}
        \right] 
        \left. \right)
\end{split}
\end{align}

Finally, we note that the variance of the inner sum is equivalent to a case where $\unittotal=1$ i.e., when a single unit is present. For this variance we can use theorem 2 of \cite{bojinov2023design}, yielding the optimization problem stated in the lemma for the single-unit case: under assumptions \cref{as:potential} \cref{as:carryover} and \cref{as:bounded} the choice of K that minimizes the variance of the estimator under adversarial selection of the potential outcomes a solution to the following subset selection problem.
$$
    \mathbb{K}^*=\min_{\mathbb{K} \subset[\steptotal]^K} \left\{4\sum_{k=0}^K (s_{k+1}-s_k)^2 + 8m(s_K-s_1) + 4m^2K - 4m^2 + 4\sum_{k=1}^{K-1}[(m-s_{k+1}+s_k)^+]^2\right\}.
$$

Therefore, minimizing the variance of $\hat{\tau}_{\lag}(W)$ is equivalent to minimizing the sum of variances for individual units, each of which follows the optimization problem from the single-unit case. Since the optimal solution for each unit is identical due to the homogeneity of the variance structure, the overall optimal design $\mathbb{K}^*$ remains the same as in the single-unit case.

The optimality of $\mathbb{Q}^* = \{\frac{1}{2},\frac{1}{2}, ...,\frac{1}{2}\}$ follows directly from Theorem 1 of \cite{bojinov2023design}, which proves it for the single-unit case. Due to \cref{as:sutva}, multi-unit designs does not affect the optimal probability assignment.
\end{proof}

\subsection{Proof of \cref{lemma:permutation}} \label{sec:proof_lemma_permutation}

\begin{proof} \noindent
To show the thesis that by randomly sampling assignment matrices $W \in \{0,1\}^{\unittotal \times \steptotal}$ following the permutation approach in the lemma we draw from a  \textit{regular} design with fixed weight $p$, we have to show that these matrices verify two things:
\begin{itemize}
    \item[(i)] for a sampled matrix $W$ all unit assignments $W_{\unitidx,\stepidx}$ are re-randomized at the same $K$ timesteps $s \in \mathbb{K}$, and
    \item[(ii)] each unit $\unitidx$ at each randomization timestep $\stepidx$ has a fixed probability $p \in (0,1)$ of being assigned to treatment.
\end{itemize}  

To show (i), we highlight that $1 < pS < S-1$ implies $ \forall \unitidx, \exists \stepidx_1, W_{\unitidx,\stepidx_1}=1$ and $\forall \unitidx \exists \stepidx_0, W_{\unitidx,\stepidx_0}=0$. 
This in turn implies that after permutation, no single $\stepidx$ is guaranteed its assignment, or equivalently $\mathbb{K}=\{1,2,..., \steptotal\}$.

We can show (ii) by observing that every row is a random permutation of $p\steptotal$ 1's and $(1-p)\steptotal$ 0's. This implies that for every \ASIN{} $\unitidx$ and position $\stepidx$, we have $p\steptotal(\steptotal-1)$ permutations that have a 1 on position $\stepidx$. 
Since there are $\steptotal!$ possible permutations $\Pr(W_{\unitidx, \stepidx} = 1) = \frac{p\steptotal (\steptotal-1)!}{\steptotal!} = \frac{p \steptotal}{\steptotal}=p$. Hence $\mathbb{Q}= \{p, p ,.., p\}$.

\end{proof}

\subsection{Proof of \cref{lemma:complement}}  \label{sec:proof_lemma_complement}

\begin{proof} \noindent
To prove that the design is balanced, we consider that any permutation $\mathbf{P}$ of a vector $\mathbf{x}$ with $\|\mathbf{x}\|_1 = p \steptotal$ still has $\|\mathbf{P(x)}\|_1 = p \steptotal$ and thus $\sum_{\stepidx=1}^\steptotal W_{\unitidx, \stepidx} = p\steptotal$. 
To prove $\sum_{\unitidx=1}^\unittotal W_{\unitidx, \stepidx} = p\unittotal$ we note that by stacking $W$ vertically with $1^{\frac{\unittotal}{2}\times \steptotal} - W$ ensures that $\sum_{\unitidx=1}^\unittotal W_{\unitidx, \stepidx} = \sum_{\unitidx=1}^\frac{\unittotal}{2} W_{\unitidx, \stepidx} + W_{\frac{\unittotal}{2}+\unitidx, \stepidx} = \sum_{\unitidx=1}^\frac{\unittotal}{2} W_{\unitidx, \stepidx} + (1-W_{\unitidx, \stepidx}) = \frac{1}{2}\unittotal = p\unittotal$.

Regularity of the design holds since regularity is a row-level property that needs to hold independently for every $\unitidx$. From \cref{lemma:permutation} we know that each row in $W$ is regular and so stacking them leads to a regular assignment as per \cref{def:regular}.
\end{proof}

\subsection{Proof of \cref{lemma:prob}}  \label{sec:proof_lemma_prob}

\begin{proof} \noindent
The number of permutations that contain $m+1$ consecutive 1's ending on a given position $\stepidx$ can be expressed as follows: we have $\binom{p\steptotal}{m+1}$ ways to pick $m+1$ 1's out of $p\steptotal$ possible 1s and those can be permuted in $(m+1)!$ ways. 
The remaining $\steptotal-(m+1)$ elements can be permuted in $(\steptotal-(m+1))!$ ways. 
Since there are $\steptotal!$ possible permutations, the probability of $m+1$ consecutive 1's is 
\[
    \frac{(m+1)!\binom{p\steptotal}{m+1}(\steptotal-(m+1))!}{\steptotal!} = 
    = 
    \frac{(p\steptotal)!}{(p\steptotal-(m+1))!(m+1)!}\frac{(\steptotal-(m+1))!(m+1)!}{\steptotal!} 
    = 
    \frac{\binom{p\steptotal}{m+1}}{\binom{\steptotal}{m+1}}.
\]
\end{proof}